\documentclass[bibyear]{aa}

\usepackage{natbib}
\usepackage{hyperref}
\usepackage{xcolor}

\usepackage{orcidlink}

\usepackage{graphicx}
\usepackage{txfonts}

\usepackage{booktabs}
\usepackage{lipsum}
\usepackage{float}

\usepackage{comment}

\newcommand{\athenapk}{{\sc AthenaPK}}

\newcommand{\bondi}{\texttt{bondi}}
\newcommand{\cool}{\texttt{cool}}
\newcommand{\turbolow}{\texttt{turb\_low}}
\newcommand{\turbohigh}{\texttt{turb\_high}}
\newcommand{\lowc}{\texttt{cca\_low}}
\newcommand{\highc}{\texttt{cca\_high}}

\hypersetup{
    colorlinks=true,
    linkcolor=blue,
    citecolor=blue,
    filecolor=blue,
    urlcolor=blue,
    menucolor=blue
}

\begin{document}

\title{{\bfseries\scshape BlackHoleWeather –} Chaotic cold accretion across the meso-scale: 
{\Large Morphology and thermodynamics}}

\titlerunning{Chaotic cold accretion from halo rain to sub-pc feeding}
\authorrunning{Filippo Barbani et al.}

   \author{Filippo Barbani \thanks{filippo.barbani@unimore.it}
        \inst{1}\orcidlink{0000-0002-1620-2577},
        Massimo Gaspari
        \inst{1}\orcidlink{0000-0003-2754-9258},
        Vieri Cammelli
        \inst{1}\orcidlink{0000-0002-2070-9047},
        Olmo Piana
        \inst{1}\orcidlink{0000-0002-1558-5289},
        Fred J. Jennings
        \inst{1}\orcidlink{0009-0000-0152-9983},
        Davide M. Brustio
        \inst{1}\orcidlink{0009-0009-7700-1910},
        Giovanni Stel
        \inst{1}\orcidlink{0009-0007-0585-9462},        
        Valeria Olivares
         \inst{2, 3}\orcidlink{0000-0001-6638-4324},
        Filippo M. Maccagni
         \inst{4, 5}\orcidlink{0000-0002-9930-1844},
        Martin Fournier
         \inst{6}\orcidlink{0009-0006-2593-1583}, 
        Francesco Tombesi
         \inst{7, 8, 9}\orcidlink{0000-0002-6562-8654},        
        Pasquale Temi
         \inst{10}\orcidlink{0000-0002-8341-342X},
        Fabrizio Fiore
         \inst{11}\orcidlink{0000-0002-4031-4157},  
        Roberto Serafinelli
         \inst{7, 12}\orcidlink{0000-0001-6638-4324}, 
        \and
        Ashkbiz Danehkar
         \inst{13}\orcidlink{0000-0003-4552-5997}
          }

   \institute{
    Department of Physics, Informatics and Mathematics, University of Modena and Reggio Emilia, I-41125 Modena, Italy
   \and
   Department of Physics, Universidad de Santiago de Chile, Santiago, Chile
    \and
    CIRAS, Universidad de Santiago de Chile, Santiago, Chile
    \and
    INAF -- Osservatorio Astronomico di Cagliari, via della Scienza 5, 09047, Selargius (CA), Italy
    \and
    Wits Centre for Astrophysics, School of Physics, University of the Witwatersrand, 2000, Johannesburg, South Africa
    \and
    Universit\"{a}t Hamburg, Hamburger Sternwarte, Gojenbergsweg 112, 21029 Hamburg, Germany
    \and
    INAF -- Astronomical Observatory of Rome, 00078 Monte Porzio Catone (Rome), Italy
    \and
    Department of Physics, University of Rome ``Tor Vergata'', 00133 Rome, Italy
    \and
    INFN -- Rome ``Tor Vergata'' Section, 00133 Rome, Italy
    \and
    NASA Ames Research Center, MS 245-6, Moffett Field, CA 94035-1000, USA
    \and
    INAF -- Astronomical Observatory of Trieste, 34143 Trieste, Italy
    \and
    Instituto de Estudios Astrof\'isicos, Facultad de Ingenier\'ia y Ciencias, Universidad Diego Portales, Avenida Ej\'ercito Libertador 441, Santiago, Chile
    \and
    Science and Technology Institute, Universities Space Research Association, Huntsville, AL 35805, USA
   }
   
   \date{\today}

 
  \abstract
{Supermassive black holes (SMBHs) self-regulate galaxies, groups, and clusters, yet the pathway that transports gas from halo scales to sub-parsec radii remains debated. In hot, stratified atmospheres, subsonic turbulence can trigger nonlinear thermal instability and a multiphase condensation cascade, producing chaotic time-variable BH `weather'. A key missing link is how the meso-scale (parsecs to kiloparsecs) connects halo rain to the nuclear inflow.}
{We study turbulence-driven condensation and chaotic cold accretion (CCA) in a group-scale halo, quantifying how the stirring level shapes multiphase morphology and thermodynamics, and how this imprints on SMBH feeding down to sub-parsec scales.}
{We ran 3D hydrodynamic hyper-zoom simulations with a GPU-accelerated code, including radiative cooling and driven subsonic turbulence in a hot intragroup halo. Two endpoint runs bracket weak and strong stirring, capturing distinct BH weather states.}
{In both regimes the atmosphere becomes thermally unstable and develops a strongly multiphase medium spanning 8-10 orders of magnitude in temperature and density. Strong stirring delays cold gas accretion and sustains an extended, filament-rich rain pattern to kiloparsec radii (`stormy' CCA), with broader thermodynamic distributions beyond the nucleus. Weak stirring triggers earlier condensation but yields a more compact clumpy rain, with most cold gas confined within 100~pc (`rainy' CCA). At micro-scales the inflow is partly mediated by a clumpy rotating torus. Despite large differences in condensed cold mass, the BH accretion rate is recurrently boosted by up to $100\times$ above the hot-mode Bondi baseline and varies only weakly between the weather regimes, indicating that feeding is regulated primarily by how efficiently multiphase structures couple to the central inflow.}
{Modest changes in turbulence are sufficient to shift the same hot halo between stormy (extended) and rainy (centralized) BH weather, providing a robust quantitative multiscale baseline for interpreting multiphase CCA and its impact on SMBH feeding.}


   \maketitle
%

\section{Introduction} \label{sec:intro}

Over cosmic time, small primordial fluctuations have grown into large-scale structures, hosting most baryons in the form of diffuse, hot plasma—the circumgalactic, intragroup, and intracluster medium \citep[CGM, IGrM, ICM;][]{WhiteFrenk1991,kravtsov2012,tumlinson2017}. These gaseous halos act as stratified atmospheres shaped by turbulence, gravity, cooling, and feedback, giving rise to a `cosmic weather' \citep{GaspariTombesi2020}. Multi-wavelength observations increasingly reveal that this hot gas undergoes multiphase condensation, forming warm, neutral, and cold components \citep[e.g.][]{fabian2008,tremblay2016,maccagnimorganti2018,maccagni2021,morganti2023}, yet the physical mechanisms behind this process and its role in galaxy evolution remain open questions.

Most galaxies host a supermassive black hole \citep[SMBH,][]{kormendy2013}, which plays a key role in their baryon cycle. Through powerful outflows and jets, SMBHs inject energy into their host halos, regulating gas cooling and heating, star formation, and the galaxy luminosity function \citep[e.g.][]{king2015,fiore2017,kartheik2025}. The ensemble of phenomena tied to a SMBH is typically known as an active galactic nucleus (AGN; \citealt{Lynden-Bell1969,mcnamara2012}). Central open problems are how gas is transported from galactic scales down to the SMBH, how angular momentum is removed, and how AGN feedback couples back to the multiphase medium.

Bridging physical processes across more than nine orders of magnitude, from the megaparsec scale cosmic environment down to the black hole gravitational radius, remains a major theoretical and numerical challenge \citep{NaabOstriker2017,VogelsbergerMarinacci2020}. Accretion is governed by the interplay of turbulence, cooling, heating, angular momentum transport, and magnetic fields \citep{balbushawley1998,YuanNarayan2014}. Connecting theory and observation therefore requires one to follow gas across this entire range of scales, with multiphase structures in hot, stratified halos providing key constraints on how cooling, turbulence, and feedback regulate inflow \citep[e.g.][]{Fabian2012}.

In the absence of turbulence and feedback, accretion would proceed as a steady, nearly spherical inflow, concentrating cold gas and star formation at the very centre \citep{Fabian1994}. In reality, ultraviolet observations of brightest cluster galaxies reveal a far more complex picture: star formation occurs in diverse morphologies—often clumpy, in knots, or filamentary—consistent with star formation triggered or enhanced by AGN jet activity \citep[e.g.][]{donahue2015, fogarty2015}. A similar connection between star-forming structures and radio jets is also supported by recent multi-wavelength observations \citep[e.g.][]{reefe2025}. These observations suggest that gas can cool and form stars at large radii ($\approx 10$–$30$ kpc) from the centre of massive halos, along extended, filamentary, and multiphase structures. Millimetre and sub-millimetre observations of molecular gas \citep[e.g.][]{russell2019,OlivaresSalome2019,castignani2025,toni2026} reveal cold gas reservoirs and, in several systems, cold filaments whose morphology closely follows that of the radio jets. Such structures are interpreted as filaments of accreting cold gas, hosting in situ star formation and likely regulated by AGN feedback. Consistent jet–cold gas alignments are also observed in high-redshift systems, where ALMA observations reveal extended cold molecular reservoirs in the CGM aligned with radio jets, supporting a close coupling between AGN activity and cold gas condensation \citep[e.g.][]{LiEmonts2021,WalterBanados2025}.

In realistic intracluster, intragroup, and circumgalactic environments the hot plasma is, however, turbulent, intermittently heated, and radiatively cooling via bremsstrahlung and metal-line emission. In this stratified medium, density perturbations can enter the nonlinear thermal instability regime: overdense fluctuations cool faster than they are mixed and restored, so they grow, decouple from the hot phase, and condense into warm and cold structures. Classical Bondi accretion \citep{bondi1952} offers a useful idealised baseline for steady, spherical, and adiabatic inflow, but fails to capture the complexity of realistic accretion in turbulent, multiphase environments. A promising new theory is chaotic cold accretion \citep[CCA;][]{gaspari2013,gaspari2015,gaspari2017,gaspari2017b}, in which turbulence and thermal instability promote the condensation of cold clouds that rain stochastically onto the SMBH. This inflow naturally produces variability and stochastic fueling, allowing the black hole to respond dynamically to the state of the surrounding atmosphere. CCA offers a coherent explanation for the observed variability, obscuration, and multi-wavelength signatures of AGN, supported by observations with ATCA, ALMA, MUSE, \textit{Chandra}, \textit{XMM-Newton}, JWST, and HST \citep[][]{MaccagniMorganti2014,voit2015,mcdonald2018,tremblay2018,waters2019,juranova2020,maccagni2021,mckinley2022,temi2022,ubertosi2023,ubertosi2025,olivaressalome2022,olivares2023,olivares2025,OSullivanRajpurohit2024,EskenasyOlivares2024}. Most simulations, however, resolve inflows only down to $\sim$1\,–\,100\,pc—several orders of magnitude above the black hole event horizon—and have primarily focused on massive cluster-scale halos. A fully consistent, multi-scale model of SMBH fueling from halo to sub-parsec scales is therefore still lacking and remains a major challenge for the field.

Direct evidence of turbulence has been observed with Hitomi in the core of the Perseus cluster, where spectral line broadening revealed a line-of-sight velocity dispersion of $\sigma_v = 164 \pm 10$ km s$^{-1}$ in gas with a temperature $kT \simeq 4$ keV \citep{hitomi2016}. More recently, XRISM Resolve observations have enabled a similar detection in the Abell 2029 cluster, measuring $\sigma_v = 169 \pm 10$ km s$^{-1}$ in a hotter atmosphere with $kT \simeq 8$ keV \citep{xrism2025}. These results build on earlier XMM--Newton/RGS constraints on line broadening in cool-core systems \citep[e.g.][]{SandersFabian2013,PintoSanders2015}. Assuming isotropic turbulence, these measurements correspond to three-dimensional Mach numbers of $\mathcal{M} \approx 0.2$–$0.3$\footnote{Three-dimensional Mach numbers are inferred assuming isotropic turbulence from the line-of-sight velocity dispersion.}, placing both systems in a weakly turbulent regime and indicating that the ICM hosts mainly subsonic turbulent motions. These direct measurements provide robust constraints on the level of subsonic turbulence in cool-core clusters, complementing indirect estimates derived from surface brightness fluctuations in high-resolution X-ray images, such as those from Chandra, which can be used to infer turbulence in the ICM \citep[][]{gasparichurazov2013,hofmann2016,khatri2016,dupourque2024,romero2025}. By analysing the power spectrum of surface-brightness and thermodynamic fluctuations, one can estimate the amplitude and injection scale of predominantly subsonic turbulence and relate these motions to the underlying density and pressure perturbations. Such measurements commonly indicate $\mathcal{M}\sim 0.1-0.5$ turbulence, spanning clusters to lower-mass groups.
As future missions like \textit{NewAthena} \citep{nandra2013} and \textit{AXIS} \citep{RussellLopez2024,KossAftab2025} are expected to extend these measurements to larger cluster samples and broader radial ranges, a more complete picture of ICM turbulence will emerge.

Numerical hydrodynamical simulations have become an indispensable tool in modern astrophysics, allowing us to model and interpret a wide range of nonlinear, multi-scale processes that are inaccessible to direct experimentation or analytical treatment. They have been successfully applied to study galaxy formation and large-scale structure \citep[e.g.][]{vogelsberger2014, dubois2014, pillepich2019, hopkins2023}, the evolution of the interstellar and circumgalactic medium \citep[e.g.][]{springel2005, marinacci2017, hopkins2018, barbani2023, barbani2025, zhang2025}, as well as BH accretion and AGN outflows \citep[e.g.][]{gaspari2012, chen2024, fournier2024, fournier2025, grete2025, sotira2025, jennings2025}. In the AGN context, high-resolution simulations are particularly effective at capturing the nonlinear interplay between cooling, turbulence, and feedback that governs the formation of multiphase gas and the stochastic fueling of SMBHs across a wide range of scales.

This work is part of the {\sc BlackHoleWeather} project (PI: Gaspari), which aims to build a unified multi-physics framework for SMBH feeding and feedback across cosmic environments, from galaxies to groups and clusters. {\sc BlackHoleWeather} combines high-resolution numerical experiments, synthetic multi-wavelength observables, and theory-guided diagnostics to connect multiphase halo weather, accretion variability, and AGN self-regulation within a single consistent picture.
Building on the controlled stratified-halo setups that established the CCA framework \citep{gaspari2013,gaspari2017}, we present the first of two complementary papers investigating the turbulence-driven condensation and accretion cascade down to sub-pc scales in a hot, stratified atmosphere representative of group-scale halos. We focus on the transitional meso-scale (parsecs to kiloparsecs) that links halo rain to the inner inflow. In this paper, we quantify the emergent multiphase morphology, phase structure and thermodynamics as a function of the turbulence regime. The companion paper \citep[][B26b hereafter]{Barbani2026b} addresses feeding variability and kinematics in the same set of simulations, characterising the temporal statistics and power spectra of accretion and the kinematical imprints via key CCA diagnostics (k-plots and $\mathcal{C}$-ratios).

To this end, we carry out a suite of idealised accretion simulations onto SMBHs, designed to isolate the competition between radiative cooling and turbulent mixing. We use \athenapk, an adaptive mesh refinement (AMR) hydrodynamical code designed for GPU architectures, to follow the gas from halo scales down to sub-parsec radii. By varying the strength of turbulence driving, we assess how this key process controls the multiphase condensation cascade from the hot to the molecular phase and the resulting mode of SMBH feeding, resolving cold, clumpy inflows near the SMBH and their role in regulating its growth.
Within {\sc BlackHoleWeather}, activities are organised into thematic work packages (WPs) spanning multiscale feeding and feedback (see main diagram in \citealt{GaspariTombesi2020}). This paper is part of WP2 (meso- to micro-scale feeding), focused on turbulence-driven condensation and accretion. Complementary efforts currently underway include simulations of micro-scale AGN jet feedback \citep[][C26a,b]{Cammelli2026a,Cammelli2026b}, simulations exploring the impact of BH spin on macro-scale feedback \citep[][P26a,b]{Piana2026a,Piana2026b}, and simulations investigating the role of dust physics in the multiphase feeding rain (Barbani et al. in prep.).

This paper is structured as follows. In Section \ref{num}, we describe the simulation setup, including the physical models, initial conditions, and numerical methods. In Section~\ref{res} we present the main results, focusing on the formation and evolution of the multiphase structures and their role in feeding the SMBH. 
In Section~\ref{sec:evol} we interpret these outcomes in terms of BH weather states and propose an evolutionary link between regimes.
In Section~\ref{sec:synthesis}, we compare our findings with previous observational and numerical studies. Finally, Section \ref{conc} summarises our conclusions and outlines the directions for future work.

\section{Numerical methods}\label{num}

\begin{figure}
\centering
\includegraphics[width=0.9\columnwidth]{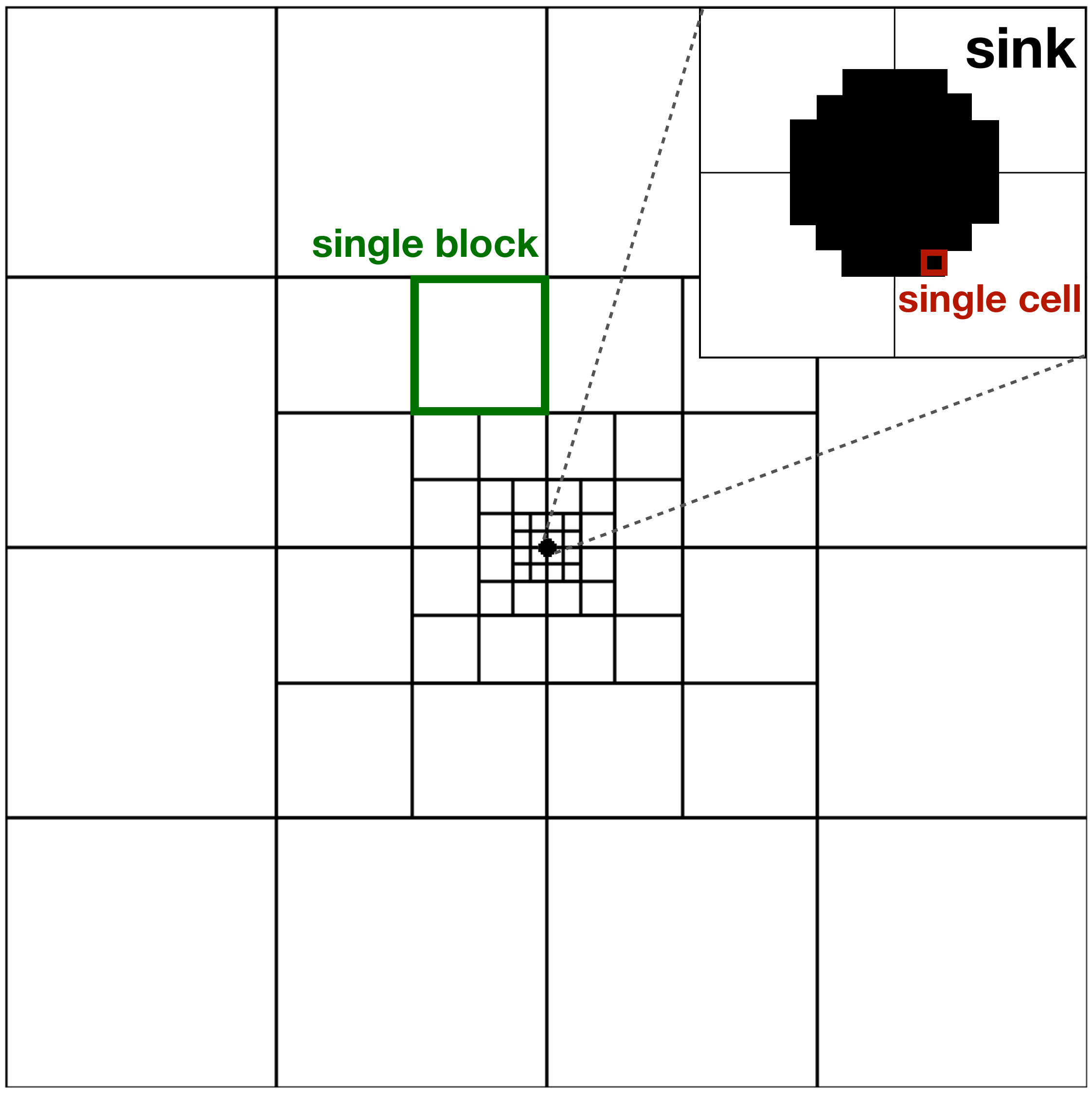}

\caption{Simulation grid and SMBH sink scheme. The domain is decomposed into blocks (shown as a green square), which are recursively refined towards the centre of the domain (each block contains $32^3$ cells). The decreasing block size increases the effective resolution in the central region, where the black hole sink is located. The black hole sink (shown in black) removes mass within a radius of four finest-level cells at each time step (an example of cell is shown as a red square).}\label{sink_scheme}

\end{figure}

The simulations performed in this work are made using the AMR code \athenapk, an open source, performance-portable magneto-hydrodynamical code, directly descending from {\sc Athena++} \citep{stone2020}, based on the block-structured adaptive mesh refinement framework {\sc Parthenon} \citep{grete23} and the performance
portability programming model {\sc Kokkos} \citep{edwards2014,trott21}. In this way, \athenapk \ is compatible with a wide range of architectures and with both CPUs and GPUs, demonstrating high efficiency and scalability up to 73,728 GPUs.

The equations of hydrodynamics that are integrated by the code in conservative and Eulerian form are as follows:

\begin{equation}
\dfrac{\partial \rho}{\partial t} + \nabla \cdot (\rho \boldsymbol{v})=0,
\end{equation}

\begin{equation}\label{eq:hydro2}
\dfrac{\partial (\rho \boldsymbol{v})}{\partial t} + \nabla \cdot (\rho \boldsymbol{v} \otimes \boldsymbol{v} + P\mathbb{I})= -\rho \nabla \Phi + \rho \boldsymbol{f}_{\text{turb}},
\end{equation}

\begin{equation}\label{eq:hydro3}
\dfrac{\partial E}{\partial t} + \nabla \cdot ( E + P ) \mathbf{v}= -\rho \mathbf{v}\cdot \nabla \Phi - \mathcal{C} + \mathcal{S}_{\text{turb}},
\end{equation}

where $\rho$ is the density, $\mathbf{v}$ is the velocity, $P$ is the pressure, $\mathbb{I}$ is the identity tensor, $\Phi$ is the total gravitational potential, $\boldsymbol{f}_{\text{turb}}$ is the turbulent driving acceleration (see Section \ref{turbulence}) and $E$ is the total energy density of the gas, defined as $E=e + \rho v^2/2$, where $e$ is the internal energy density. $\mathcal{C}=n^2 \Lambda(T)$ is the cooling rate, which depends on the cooling function $\Lambda(T)$ (see Section \ref{radcool}) and the number density of the gas $n$, and $\mathcal{S}_{\text{turb}}= \rho \boldsymbol{v} \cdot \boldsymbol{f}_{\text{turb}}$.

The code employs a finite-volume \citet{godunov1959} scheme with second-order accuracy in both space and time. Given the highly multiphase chaotic medium, for stability, we use the piecewise linear method \citep[][]{vanLeer1979} for spatial reconstruction and a second-order Runge–Kutta scheme \citep{butcher2008} for time integration. Fluxes at cell interfaces are calculated by solving the Riemann problem using the approximate Harten-Lax-van Leer contact (HLLC) solver \citep{toro1994}, which accurately captures contact and shear discontinuities. Given the complexity of astrophysical phenomena, we applied a first-order flux correction to ensure numerical stability and accuracy near discontinuities \citep{bruggen2023}. In gas cells where non-physical values, such as negative densities or temperatures, arise, fluxes are recalculated using a forward Euler time integration, first-order (piecewise constant) reconstruction, and the Local Lax–Friedrichs Riemann solver. This fallback approach ensures physically meaningful solutions without the need to impose artificial density or temperature floors.

\begin{figure*}
\centering
\includegraphics[width=\textwidth]{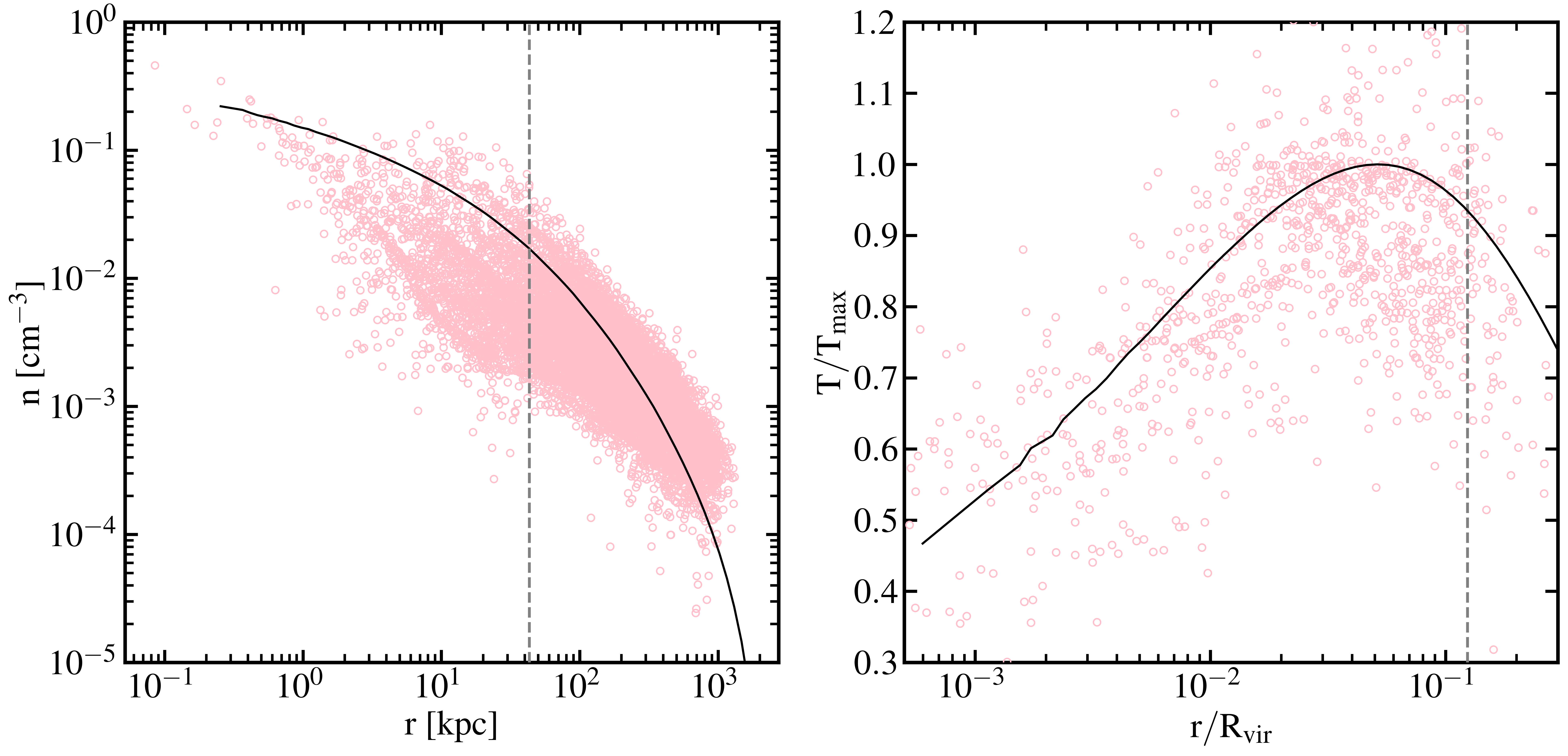} 

\caption{Thermodynamic properties of the simulated galaxy group at $t=0$ Myr (black lines). Left: radial gas number density profile compared with observed electron densities from the ACCEPT database \citep[][pink circles]{cavagnolo2009}. Right: normalised temperature profile ($T/T_{\rm max}$) as a function of $r/R_{\rm vir}$, compared with early-type galaxies from the \textit{Chandra} Galaxy Atlas \citep[][pink circles]{kim2020}. The ICs are consistent with observed systems, and the temperature profile follows a nearly universal shape with a broad peak around $r \simeq 0.04\,R_{\text{vir}}$. The vertical dashed line marks the maximum radius contained in the computational domain ($r_{\rm max}=43.3$ kpc, corresponding to half of the box diagonal).}\label{profile_IGM}

\end{figure*}

\subsection{Grid structure}

The simulation is performed using a static mesh refinement (SMR) setup. The computational domain is organised as a hierarchy of nested Cartesian grid blocks with increasing resolution towards the centre of the domain, where the highest spatial accuracy is required. Each grid block contains $32^3$ cells, and refinement proceeds by subdividing parent blocks into child blocks with twice the resolution in each spatial direction. The refinement hierarchy is static, defined at initialisation, and remains fixed throughout the simulation. It is centred on the origin and extends radially outwards, with higher refinement levels covering progressively smaller volumes, while coarser grids are employed in the outer halo to reduce computational cost and still capture the large-scale gas dynamics (see Figure~\ref{sink_scheme}).
The root grid spans the entire simulation volume and contains $128^3$ cells, with a total of $\ell = 12$ refinement levels. The cell size at refinement level $\ell$ is given by
\begin{equation}
\Delta x_{\ell} = \frac{L}{N_0 \cdot 2^{\ell}},
\end{equation}
where $L = 50$ kpc is the box size and $N_0 = 128$ is the number of cells per side of the root grid. This configuration yields a maximum spatial resolution of $\Delta x \simeq 0.1$ pc within the central 3.2~pc region.
Consistency across refinement levels is ensured through appropriate prolongation and restriction operations at all refinement boundaries. This SMR configuration enables accurate modelling of multiscale processes such as gas cooling, inflow, and feedback-driven turbulence in the central region of the halo.

\subsection{SMBH sink}

Following \citet{gaspari2013} setup, to represent the presence of a central SMBH, we introduce a sink region at the centre of the computational domain. The sink is implemented as a fixed spherical region of radius $r_{\rm sink} = 4\Delta x_{\text{min}}= 0.4$ pc, centred at the origin, and acts as an effective inner boundary condition for gas accretion.

At each timestep, gas within $r_{\rm sink}$ is removed from the simulation by resetting its thermodynamic and kinematic properties to $\rho_{\rm sink} = 10^{-30} \ \mathrm{g \ cm^{-3}}$, $T_{\rm sink} = 1 \ \mathrm{K}$, and $v_{\rm sink} = 0$ km s$^{-1}$. This prescription avoids nonphysical gas pile-up at the resolution limit of the simulation and maintains numerical stability. Thanks to the sub-parsec spatial resolution achieved in the central region, the simulations resolve gas dynamics well within the Bondi radius of a group-scale SMBH (i.e. $r_B\sim$1--30 pc, see Section \ref{ICs}). However, the physical processes occurring at event-horizon scales remain unresolved. The sink therefore provides a physically motivated approximation for the unresolved accretion flow, absorbing gas that would otherwise be expected to feed the SMBH. The gravitational softening length $\epsilon$ (Plummer softening, i.e. $\Phi = -GM/\sqrt{r^2+\epsilon^2}$) is chosen to be smaller than $r_{\rm sink}$, ensuring that gravitational forces are well resolved within the sink region. The SMBH mass is fixed and does not evolve during the simulation. The total mass that would be accreted over the simulated time span is negligible compared to the assumed black hole mass, and does not affect the gas dynamics or the gravitational potential in the central region.

\subsection{Initial conditions}\label{ICs}

The initial conditions (ICs) consist of a gaseous halo in hydrostatic equilibrium inserted in a static gravitational potential.

Since galaxy groups are the fundamental building blocks of the cosmic web and host the bulk of present-day baryons in halos, we focus on a representative, intermediate-mass system rather than an extreme, massive cluster. This also helps the computational costs where the cooling time is significantly shorter (i.e.~$t_{\text{cool}}\approx 10-20$ Myr in the central kiloparsec of our simulations).

Gravity is modelled using a static gravitational potential that includes contributions from a dark matter halo, a central dominant group galaxy, and a supermassive black hole (see also \citealt{fournier2024}). The gravitational acceleration is defined as $\mathbf{g}=-\nabla \Phi$. For the dark matter halo, we assume a Navarro–Frenk–White (NFW) density profile \citep*{nfw1997}, for which the radial component of the gravitational acceleration is

\begin{equation}
g_{\text{NFW}}(r) = 
  \frac{G}{r^2}
  \frac{\text{M}_{\text{NFW}}  \left [ \ln{\left(1 + \frac{r}{R_{\text{NFW}}} \right )} - \frac{r}{r+R_{\text{NFW}}} \right ]}
        { \ln{\left(1 + c_{\text{NFW}}\right)} - \frac{ c_{\text{NFW}}}{1 + c_{\text{NFW}}} },
\end{equation}
where $\text{M}_{\text{NFW}}$ is the virial mass of the halo, $R_{\text{NFW}}$ is the scale radius, and $c_{\text{NFW}}$ is the halo concentration parameter.

The scale radius $R_{\text{NFW}}$ is related to the total halo mass and the characteristic density $\rho_s$, and is defined by the following:

\begin{equation}
R_{\text{NFW}} = \left ( \frac{\text{M}_{\text{NFW}}}{ 4 \pi \rho_s \left [ \ln{\left ( 1 + c_{\text{NFW}} \right )} - c_{\text{NFW}}/\left(1 + c_{\text{NFW}} \right ) \right ] }\right )^{1/3},
\end{equation}

which ensures that the mass enclosed within $R_{200}=c_{\text{NFW}}R_{\text{NFW}}$ matches $\text{M}_{\text{NFW}}$. The characteristic density $\rho_s$ determines the amplitude of the NFW density profile and is calculated as

\begin{equation}
\rho_s = \frac{200}{3} \rho_{\text{crit}} \frac{c_{\text{NFW}}^3}{\ln{\left ( 1 + c_{\text{NFW}} \right )} - c_{\text{NFW}}/\left(1 + c_{\text{NFW}} \right )},
\end{equation}

where the critical density of the Universe, $\rho_{\text{crit}}$, sets the normalisation, and is defined as

$$
\rho_{\text{crit}} = \frac{3\text{H}_0^2}{8\pi G},
$$

with H$_0$ being the Hubble parameter. This formulation allows the NFW profile to be fully specified by the two parameters $\text{M}_{\text{NFW}}$ and $c_{\text{NFW}}$, which are typically derived from cosmological simulations or observational constraints.

The central dominant elliptical galaxy (cD) is modelled with a Hernquist profile \citep{hernquist1990}, which gives a gravitational acceleration $g(r)$ of
\begin{equation}
 g_{\text{cD}}(r) = G \frac{ \text{M}_{\text{cD}} }{R_{\text{cD}}^2} \frac{1}{\left( 1 + \frac{r}{R_{\text{cD}}}\right)^2},
\end{equation}
where $R_{\text{cD}}= 10$ kpc and $M_{\text{cD}}=1.4\times10^{11}$ M$_{\odot}$, similar to NGC 5044, where recurrent evidence for CCA has been found \citep[e.g.][]{temi2022, Rajpurohit2025}.

The gravity of the SMBH is derived from its mass as a point-like source as

\begin{equation}
g_{\text{SMBH}}(r)=\frac{GM_{\bullet}}{r^2},
\end{equation}
where the SMBH mass is set to $M_{\bullet}=2.8\times10^8$ M$_{\odot}$, representative of a central black hole in a
low-redshift galaxy group and consistent with the seeding prescriptions adopted in semi-analytic and cosmological models of black hole formation \citep[e.g.][]{PianaDayal2021,CammelliMonaco2025,PianaPu2025}. Therefore, the total gravitational acceleration is given by $g_{\text{tot}} = g_{\text{NFW}} + g_{\text{cD}} + g_{\text{SMBH}}$.

The entropy profile is defined following a power law as in the ACCEPT cluster database \citep{cavagnolo2009}, with values rescaled to represent a galaxy group,

\begin{equation}
 K(r) = K_{0} + K_{100} \left ( r/ 100 \text{ kpc} \right )^{\alpha_K},
\end{equation}
where $K_0$ is the central entropy, $K_{100}$ is the entropy at $r=100$ kpc, and $\alpha_K$ is the slope of the power law.

Knowing the gravitational acceleration $g(r)$ and the entropy $K(r)$ profiles, we can derive the density profile $\rho(r)$, assuming hydrostatic equilibrium,
\begin{equation}
\frac{1}{\rho} \frac{\text{d}P}{\text{d}r} = -g(r),
\end{equation}
which can be rewritten in terms of entropy as
\begin{equation}
\frac{\text{d}}{\text{d}r}[K(r)\rho(r)^{\gamma}] = -\rho(r)g(r).
\end{equation}

\begin{figure}
\centering
\includegraphics[width=0.9\columnwidth]{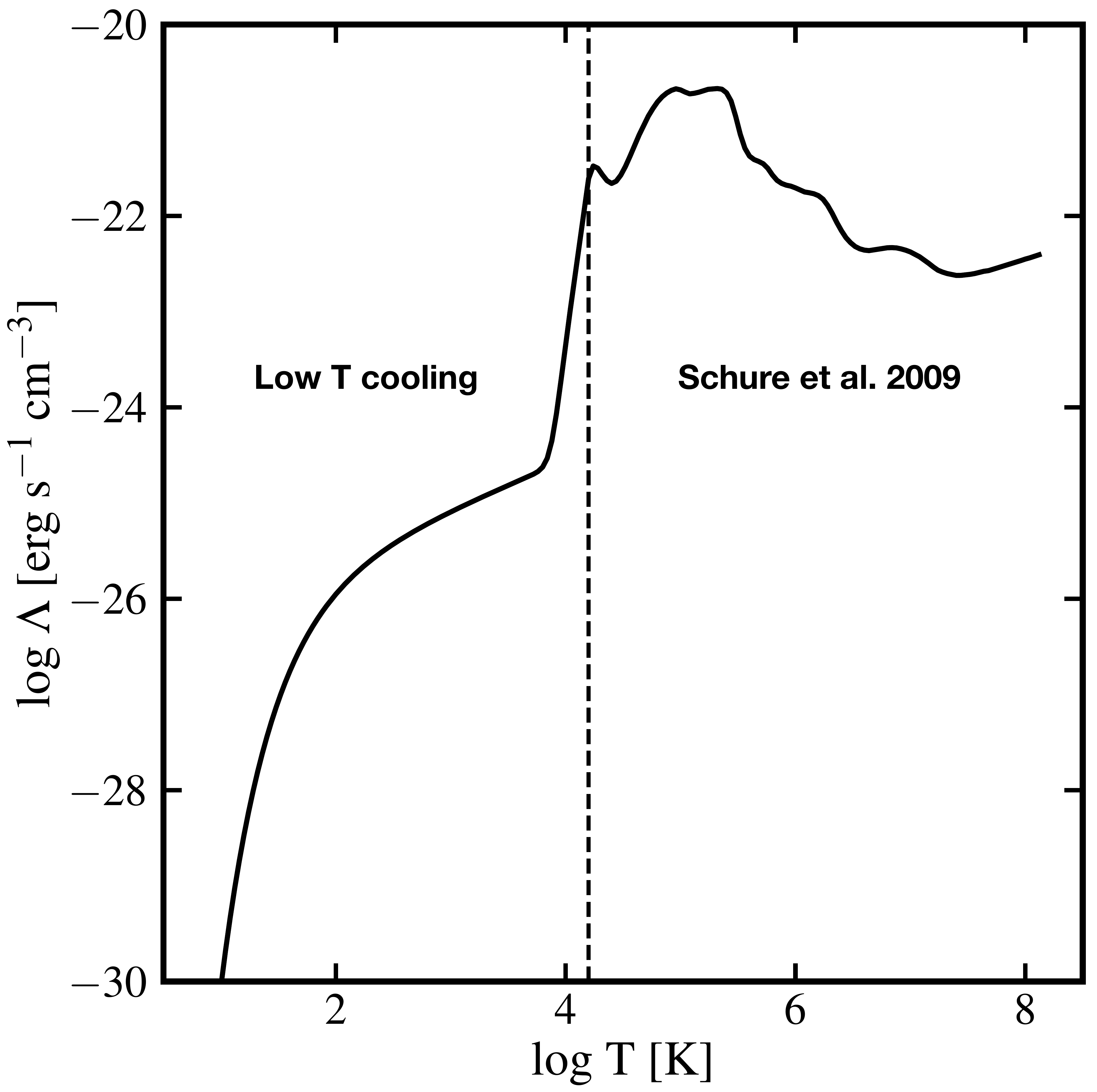}

\caption{Radiative cooling function at solar metallicity used in our simulations. The curve shows $\Lambda(T)$ as a function of temperature, with tabulated values from \citet{schure2009} for $T \ge 10^{4.2}\,\mathrm{K}$ and the analytic fit given by Eq.~(\ref{lambda}) for $T < 10^{4.2}\,\mathrm{K}$. 
Cooling function includes metal-line, recombination, bremsstrahlung and low temperature cooling; it is tabulated on a uniform grid in $\log T$ and interpolated during the simulation to compute radiative losses.}\label{cooling_function}

\end{figure}

\renewcommand{\arraystretch}{1.2}
\begin{table*}
        \centering
        \caption{Structural parameters of the galaxy group simulated in this work.}\label{tabella_param}
        \begin{tabular}[t]{ccccccccccccc}
        
                \hline
        \hline
                M$_{\text{NFW}}$&$R_{\text{NFW}}$&$c_{\text{NFW}}$&H$_0$ &M$_{\text{cD}}$&$R_{\text{cD}}$&M$_\bullet$& K$_0$ &K$_{100}$ & $\alpha_K$\\

                [M$_{\odot}$]&[kpc]& &[$\mathrm{km\,s^{-1}\,Mpc^{-1}}$] &[M$_{\odot}$]&[kpc]&[M$_{\odot}$] & [keV cm$^2$] & [keV cm$^2$] \\[0.15cm]
                \hline
                $1.5\times 10^{13}$ & 36.46 & 5 & $70$ & $1.4\times 10^{11}$ & $10$ & $2.8\times10^8$ & 0.34 & $71$ & 0.5\\
                
                \hline
        \end{tabular}
    \tablefoot{From left to right: Dark matter halo mass (M$_\text{NFW}$); dark matter halo scale length ($R_{\text{NFW}}$); dark matter halo concentration parameter ($c_{\text{NFW}}$); Hubble constant (H$_0$); cD galaxy mass (M$_\text{cD}$); cD galaxy scale length ($R_{\text{cD}}$); SMBH mass (M$_\bullet$); central entropy (K$_0$); entropy at 100 kpc (K$_{100}$); entropy power law slope ($\alpha_K$).}
\end{table*}

To derive the density, we first need to close the system defining the electron density $n_e(r_{\text{fix}})$ at a fixed radius $r_{\text{fix}}=10$ kpc. The derived density and temperature profiles are shown in Figure~\ref{profile_IGM}. 
The pink circles correspond to observations of electron density in galaxy groups taken from the \textsc{ACCEPT} catalogue, while the solid black lines represent the profiles derived from our simulation. The simulated density profile lies within the observed scatter and broadly reproduces the characteristic slope and normalisation of the intragroup and intracluster medium over more than two orders of magnitude in radius. In the right panel, the temperature profile of our ICs also agrees well with observational trends, following the universal shape identified by \citet{kim2020} from Chandra observations of early-type galaxies, when both the temperature and radius are normalised by the peak temperature ($T_{\mathrm{peak}}$) and the virial radius ($R_{\mathrm{vir}}$), respectively. All parameters relevant for the definition of our ICs are listed in Table \ref{tabella_param}.

\subsection{Radiative cooling}\label{radcool}

Radiative processes are the main cause of gas cooling in the Universe and are a fundamental process for the formation and evolution of galaxies. The cooling function $\Lambda$ is shown in Figure \ref{cooling_function}. For $T \ge 10^{4.2}\,\mathrm{K}$, we adopt the high-temperature cooling curve of \citet{schure2009} with solar metallicity. Their model assumes an optically thin plasma in collisional ionisation equilibrium and includes radiative losses from a very extensive set of metal-line transitions in the ultraviolet and soft X-ray bands, together with free--free and recombination continuum emission. The cooling function accounts for thousands of transitions from the main astrophysical elements (H, He, C, N, O, Ne, Mg, Si, S, Ar, Ca, Fe, Ni), providing a physically consistent treatment of radiative losses in warm and hot gas.

For $T<10^{4.2}$ K, we use the prescription from \citet{gaspari2017} with a volumetric cooling rate defined as $n_H^2 \Lambda_{\text{cold}}$

\begin{align}\label{lambda}
\Lambda_{\text{cold}} &= 2 \times 10^{-19} 
   \exp[-1.184\times 10^5/(T+10^3)] \notag \\
&\quad + 2.8 \times 10^{-27}\sqrt{T}\,\exp[-92/T],
\end{align}
which incorporates crucial low temperature processes such as atomic line cooling, rotovibrational line cooling, and molecular collisions with dust grains. This implementation is crucial to realistically generate cold (10-100 K) gas.

We integrated the cooling source term, $\mathcal{L}$, with an exact implicit scheme \citep{Townsend2009,gaspari2012}, which provides an analytically exact update of the thermal energy over each timestep. This method is unconditionally stable, preserves positive internal energies, and eliminates the stringent timestep constraints normally imposed by stiff cooling functions, making it well suited for simulations where gas can cool rapidly over many orders of magnitude in temperature.

\subsection{Turbulence driving}\label{turbulence}

No explicit AGN feedback through jets and/or winds is included in this setup, as our primary focus is on the dynamics of the pure gas inflow transitioning from macro- to micro-scales. Nevertheless, we include gentle, large-scale AGN-motivated turbulence to capture the dynamical impact of SMBH activity on the surrounding medium. This approach provides a controlled framework for isolating inflow and cooling processes in the central regions of group-scale halos.

To sustain turbulence, we drive the gas with a stochastic acceleration field
constructed from a superposition of random Fourier modes and evolved as an
Ornstein--Uhlenbeck (OU) process \citep[][]{schmidt2009,grete2018,grete2025},
following the original CCA setup \citep{gaspari2013,gaspari2017}, in which
continuous, predominantly subsonic stirring of a stratified atmosphere seeds a
turbulent cascade and fosters nonlinear multiphase condensation.

The forcing peaks at a characteristic dimensionless mode number
$n_{\rm peak}$, corresponding to an injection scale
$L_{\rm inj}=L_{\rm box}/n_{\rm peak}$. The associated physical
wavenumber is $k_{\rm peak}=2\pi n_{\rm peak}/L_{\rm box}$. The OU process is characterised by a correlation time $t_{\rm corr}$, while the solenoidal versus compressive composition of the forcing is controlled by a weighting parameter $\zeta$, such that $\zeta=1$ yields purely solenoidal forcing and $\zeta=0$ purely compressive forcing.

In Fourier space, the OU update of each individual acceleration mode reads
\begin{equation}
\hat{a}_i(\boldsymbol{k}, t+\Delta t)
= c_{\rm drift}\,\hat{a}_i(\boldsymbol{k}, t)
+ \sqrt{1 - c_{\rm drift}^2}\,
P_{\rm a}(k)\,
\mathcal{P}_{ij}\,
\mathcal{N}_j,
\label{eq:driving}
\end{equation}
where $\boldsymbol{k}$ is the Fourier wavevector, $c_{\rm drift}=\exp(-\Delta t/t_{\rm corr})$ sets the temporal correlation,
$\mathcal{N}_j$ are complex random numbers with zero mean and modulus smaller than
unity, and the target spectrum peaks at $k_{\rm peak}$ with
\begin{equation}
P_{\rm a}(k)
= \tilde{k}^2\,(2-\tilde{k}^2)\,
\Theta(2-\tilde{k}^2),
\end{equation}
where $\tilde{k}\equiv k/k_{\rm peak}$ and $\Theta$ denotes the Heaviside step function, which confines the forcing to a
narrow shell around $k_{\rm peak}$.

\begin{table}
    \centering
    \caption{Summary of the simulation suite.}
    \label{models}
    \begin{tabular}{lccc}
        \hline
        \hline
        \textbf{Simulation Name} & cooling & turbulence \\
        \hline
        \bondi  & no & no \\
        \cool & yes & no \\
        \turbolow  & no & weak \\
        \lowc  & yes & weak \\
        \turbohigh    & no & strong \\
        \highc    & yes & strong \\
        \hline
    \end{tabular}
    \tablefoot{Summary of the simulation suite. Each run is labelled by its short name and characterised by the physical processes included: radiative cooling and turbulence driving.}
\end{table}

\begin{table}
    \centering
    \caption{Turbulence driving parameters.}
    \label{tab:turbulence}
    \begin{tabular}{lccccc}
        \hline
        \hline
        \textbf{turbulence} & $a_{\mathrm{rms}}$ & $n_{\mathrm{peak}}$ & $N_{\mathrm{modes}}$ & $\zeta$ & $t_{\mathrm{corr}}$ \\
        & [$10^{-8}\mathrm{cm\,s^{-2}}$] &  &  &  & [Myr] \\
        \hline
        weak  & 0.62 & 4 & 64 & 1 & 30 \\
        strong & 1.55 & 4 & 64 & 1 & 30 \\
        \hline
    \end{tabular}
    \tablefoot{
        Parameters of the stochastic turbulence-driving module. The acceleration field has an RMS amplitude $a_{\mathrm{rms}}$ and is injected with a solenoidal fraction $\zeta=1$, corresponding to purely incompressible driving. The power spectrum peaks at the dimensionless mode number $n_{\mathrm{peak}}=4$, with $N_{\mathrm{modes}}=64$ random Fourier modes refreshed every $t_{\mathrm{corr}}=30~\mathrm{Myr}$.  
        The two sets of simulations differ only in the driving strength: the weak-turbulence runs (\turbolow, \lowc) and the strong-turbulence runs (\turbohigh, \highc).
    }
\end{table}

The solenoidal or compressive nature of the forcing is imposed through the
projection tensor which implements the Helmholtz decomposition
\begin{equation}
\mathcal{P}_{ij}
= \zeta\,\delta_{ij}
+ (1-2\zeta)\,\frac{k_i k_j}{|\boldsymbol{k}|^2},
\label{eq:driving2}
\end{equation}
where $\delta_{ij}$ is the Kronecker delta. Eq.~(\ref{eq:driving}) describes the OU evolution of each individual Fourier
coefficient. The real-space turbulent acceleration field is then constructed by
summing a finite number $N_{\rm modes}$ of such modes, whose wave-vectors
$\boldsymbol{k}_m$ are randomly drawn within a narrow shell around
$k_{\rm peak}$:
\begin{equation}
\boldsymbol{a}(\boldsymbol{x},t)
=
\Re\!\left[
\sum_{m=1}^{N_{\rm modes}}
\hat{a}(\boldsymbol{k}_m,t)\,
e^{i\boldsymbol{k}_m\cdot\boldsymbol{x}}
\right],
\label{eq:Nmodes}
\end{equation}
where $\Re$ denotes the real part of the complex Fourier sum. The turbulent forcing term entering the hydrodynamic equations (\ref{eq:hydro2} and \ref{eq:hydro3}) is defined as $\boldsymbol{f}_{\rm turb}(\boldsymbol{x},t)
\equiv
\boldsymbol{a}(\boldsymbol{x},t)$, where $\boldsymbol{a}$ is the real-space stochastic acceleration field obtained
from the Fourier-space OU driving described above. The overall forcing strength
is controlled by the prescribed root-mean-square (RMS) acceleration amplitude,
\begin{equation}
a_{\rm rms}
\equiv
\left\langle
|\boldsymbol{a}(\boldsymbol{x},t)|^2
\right\rangle^{1/2},
\end{equation}
where angle brackets denote a volume average over the computational domain. At
each timestep, the acceleration field is renormalised to match the prescribed
value of $a_{\rm rms}$ and to ensure zero net momentum injection.

This method injects kinetic energy at large scales in a controlled and
statistically stationary way. The resulting turbulent velocity dispersion and
energy injection rate are determined by the combination of $a_{\rm rms}$,
$L_{\rm inj}$, and $t_{\rm corr}$. The parameters used for our simulations are listed in Table \ref{tab:turbulence}. Energy is injected solenoidally at an intermediate spatial scale, $L_{\rm inj} = 12.5\,\mathrm{kpc}$, which is neither purely small-scale nor fully large-scale. As such, this choice is not uniquely associated with a single physical mechanism, but is broadly consistent with turbulence driven by AGN feedback, sloshing and/or minor mergers \citep[e.g.][]{gaspari2012,vazza2012}. This intermediate-scale forcing initiates a turbulent cascade in which kinetic energy is transferred non-linearly from the injection scale to smaller scales, where it is eventually dissipated. This approach allows the system to self-consistently develop a turbulent spectrum characteristic of galaxy groups and clusters.

\subsection{Simulation suite}

All simulations were performed with the \athenapk\ code using SMR. The simulation box has a size of $L_{\text{box}}=50$ kpc, with 12 nested levels of refinement reaching a finest resolution of $\simeq0.1~\mathrm{pc}$ within the inner region. All the runs start from identical initial density and temperature profiles (see Figure \ref{profile_IGM}) representing a static hot gaseous halo in hydrostatic equilibrium.

The simulation suite is summarised in Table~\ref{models}. We progressively add physical processes to isolate their effects on the gas dynamics. The \bondi\ run is obtained with an adiabatic simulation, with both driven turbulence and radiative cooling not active, and we use it as a benchmark to compare to the other simulations. The \turbolow\ and \turbohigh\ runs include stochastic solenoidal driven turbulence with root-mean-square accelerations of $a_{\mathrm{rms}}= 6.2\times10^{-9}~\mathrm{cm\,s^{-2}}$ and $a_{\mathrm{rms}}\simeq1.6\times10^{-8}~\mathrm{cm\,s^{-2}}$, but without radiative cooling. These two cases isolate the dynamical impact of different levels of subsonic turbulent stirring in a stratified hot atmosphere. Furthermore, we carry out two main simulations to investigate how the strength of turbulence influences the development of CCA. Both runs include radiative cooling but differ in their level of turbulence during the simulation time. We test two extremal weather regimes: a weak turbulence case (\lowc) and a strong turbulence case (\highc), characterised by typical velocity dispersions of $\sigma_v \simeq 60$–$90~\mathrm{km~s^{-1}}$ and $\sigma_v \simeq 210$–$230~\mathrm{km~s^{-1}}$, respectively. Radiative cooling is activated after an initial evolution of 35 Myr, allowing turbulence to evolve towards our target velocity dispersions before condensation begins.

In our simulations, the sound speed of the IGrM in the central region is $c_s \approx 500$–$600~\mathrm{km~s^{-1}}$, corresponding to average turbulence (3D) Mach numbers of $\mathcal{M} \sim 0.15$ in \lowc\ and $\mathcal{M} \sim 0.4$ in \highc. Both cases therefore exhibit subsonic turbulence but differ by a factor of $\sim$3, bracketing direct and indirect observational constraints (e.g.~XRISM and fluctuations power spectra; see Section \ref{sec:intro}). Note that galaxy groups can reach higher Mach numbers due to the lower sound speed or X-ray halo temperature \citep[e.g.][]{hofmann2016}. 
These bracketing setups provide a controlled framework to study how turbulence and cooling jointly regulate multiphase condensation and AGN feeding under significantly different weather conditions. Throughout the paper, we adopt a `BH weather'\footnote{In analogy with terrestrial weather, a chaotic and complex system, characterised by a high sensitivity to ICs.} terminology to qualitatively label different thermodynamical and dynamical states of the IGrM, using `sunny', `rainy', and `stormy' conditions (see Figure~\ref{accretionlowc} for a quantitative illustration of these regimes and Figure~\ref{scheme} for the conceptual framework) to denote, respectively, hot turbulence-dominated atmospheres and two different states of CCA.

\begin{figure*}
\centering
\includegraphics[width=1\textwidth]{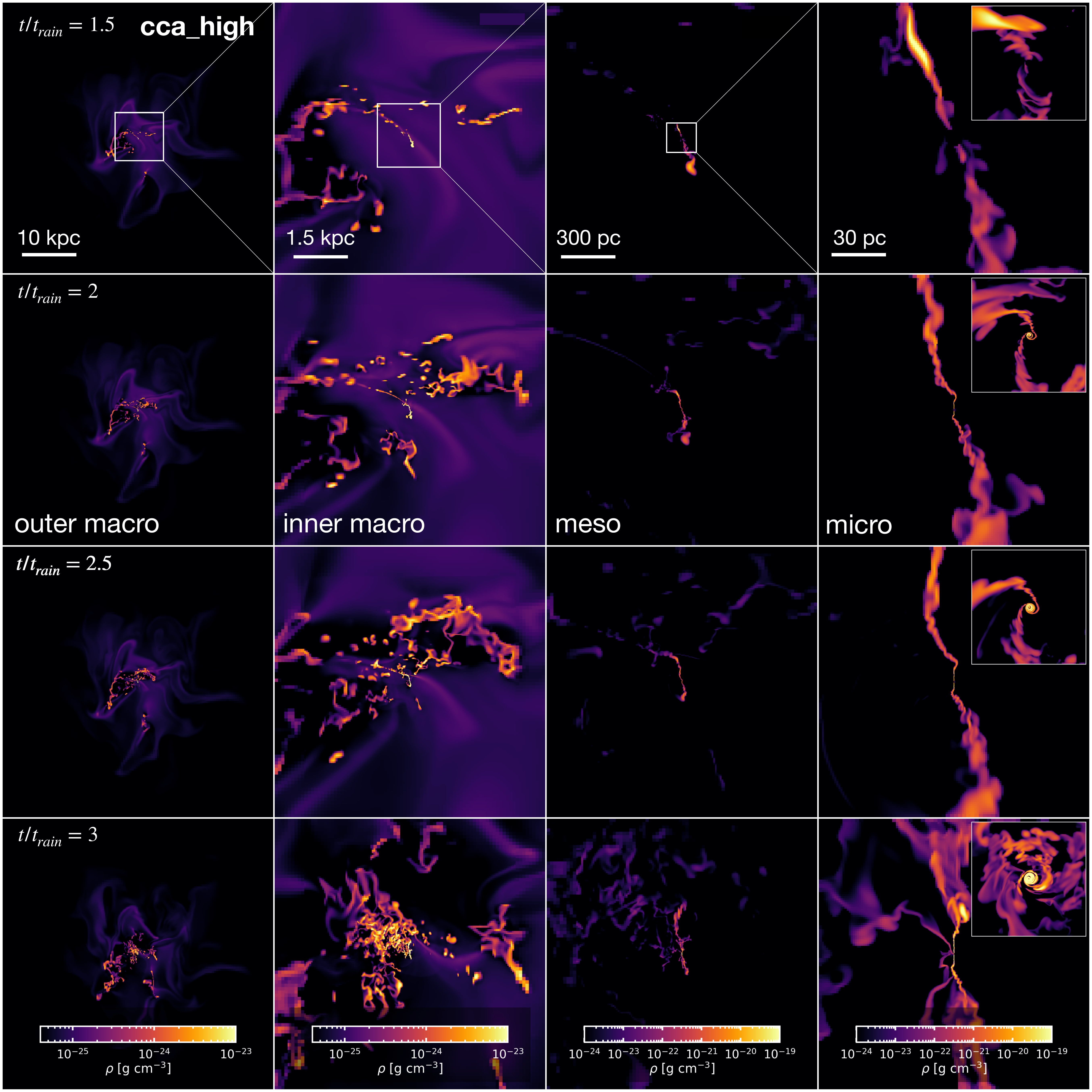}

\caption{Gas density slices of the central 50 kpc region in the \highc\ simulation.
Rows show the time evolution from $t/t_{\mathrm{rain}} = 1.5$ (top) to $t/t_{\mathrm{rain}} = 3$ (bottom), while within each row the panels from left to right present a progressive zoom-in from the halo scale (macro) to the filamentary condensation region (meso) and finally to the innermost clumps (micro). The sequence reveals the characteristic morphology of CCA: a complex network of cold, dense filaments condensing out of the turbulent hot halo and converging towards the galactic centre. Panels at different spatial scales adopt independent logarithmic colour ranges to optimise contrast and highlight the gas substructures; colour scales are thus not directly comparable across panels. The top right corner of each panel in the last column shows a zoom-in of the innermost rotating structure.}\label{density_proj_highc}
\end{figure*}

\begin{figure*}
\centering
\includegraphics[width=1\textwidth]{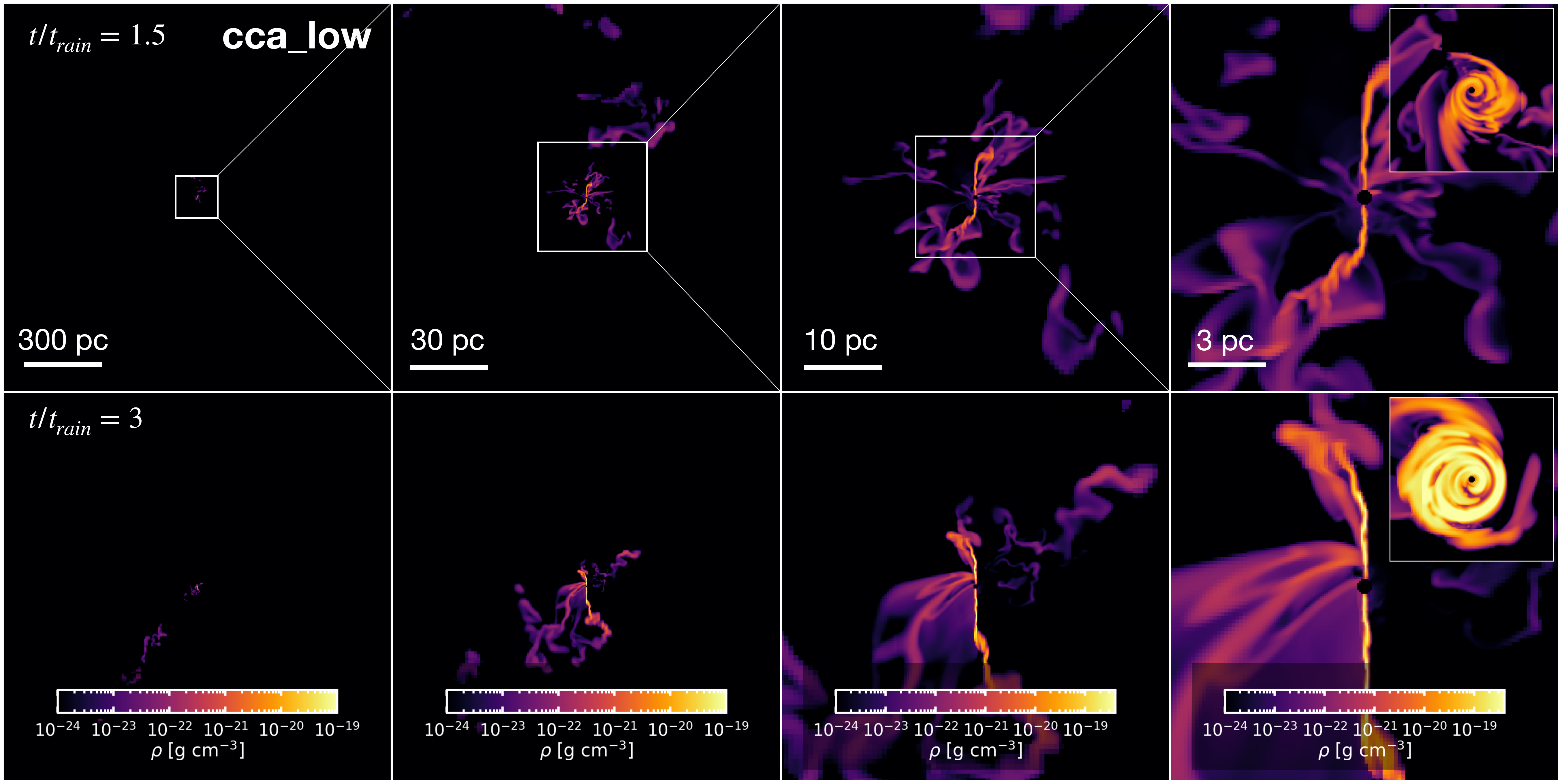}
\caption{Gas density slices of the central region in the \lowc\ simulation.
Rows show the time evolution from $t/t_{\mathrm{rain}} = 1.5$ (top) to $t/t_{\mathrm{rain}} = 3$ (bottom), while within each row the panels from left to right present a zoom-in from the inner kiloparsec to 15 pc around the SMBH. At both epochs, the system hosts a dense, clumpy cold core on parsec scales, continuously fed by smoother filamentary inflows from larger radii. The top right corner of each panel in the last column shows a zoom-in of the innermost rotating structure. Cooling dominates over turbulent mixing, allowing cold gas to accumulate and recycle within a long-lived multiphase region where clouds and filaments move along intersecting, colliding, and shearing trajectories. The morphology remains irregular and fragmented, illustrating the CCA mode also in a weakly turbulent galactic atmosphere.}\label{density_proj_lowc}
\end{figure*}

\section{Results}\label{res}

We now present the key results of our analysis, focusing on how turbulence affects gas condensation and accretion onto the SMBH. Our aim is to investigate how multiphase structure and inflow dynamics differ between the simulations with strong (\highc) and weak (\lowc) turbulence. Although we treat them as distinct setups, \highc\ and \lowc\ can also be interpreted as two possible evolutionary stages of the same system (see Section \ref{sec:evol}). Because stronger stirring enhances mixing and provides additional non-thermal support, the hot atmosphere in \highc\ requires a longer time to cool and condense than in \lowc, so the two simulations evolve on intrinsically different physical timescales. For this reason, we normalise the time by the raining time $t_{\text{rain}}$, defined as the moment when the first cold ($T\approx 10^4$ K) raining gas appears. This normalisation allows for a direct comparison of their evolution relative to their respective raining cycles. $t_{\rm rain}$ is measured starting from the moment when radiative cooling is switched on. Each simulation is evolved up to $t/t_{\text{rain}} = 3$, corresponding to $t_{\text{rain}}=30$ Myr for \highc\ and $t_{\text{rain}}=7$ Myr for \lowc\ (tied to the very inner cooling times).

\begin{table}[h!]
    \centering
    \caption{Thermal phase classification used in this work.}
    \label{tab:thermal_phases}
    \begin{tabular}{lc}
        \hline
        \hline
        \textbf{Phase} & \textbf{Temperature range} \\
        \hline
        Cold molecular (radio) & $T < 2\times10^{2}\,\mathrm{K}$ \\
        Cold (optical)           & $2\times10^{2} \le T < 1.6\times10^{4}\,\mathrm{K}$ \\
        Warm (UV)           & $1.6\times10^{4} \le T < 1.16\times10^{6}\,\mathrm{K}$ \\
        Hot (soft X-ray) & $1.16\times10^{6} \le T < 5.8\times10^{6}\,\mathrm{K}$ \\
        Hot (hard X-ray) & $T \ge 5.8\times10^{6}\,\mathrm{K}$ \\
        \hline
    \end{tabular}
\tablefoot{The temperature intervals define the thermal phase bins used throughout the paper. The labels in parentheses indicate the approximate observational band in which each phase predominantly emits.}

\end{table}

For each analysis, we divided the gas into five thermal phases corresponding to distinct temperature ranges and observational bands: cold molecular (radio), cold (optical), warm (UV), and hot gas further separated into soft X-ray and hard X-ray components. The temperature boundaries for each phase are listed in Table~\ref{tab:thermal_phases}. To investigate how the multiphase components form and evolve across scale (cf.~the {\sc BlackHoleWeather} diagram in \citealt{GaspariTombesi2020}), we partitioned the computational domain into four radial regimes: micro ($r \leq 0.1\,\mathrm{kpc}$), meso ($0.1 < r \leq 1\,\mathrm{kpc}$), inner macro ($1 < r \leq 10\,\mathrm{kpc}$), and outer macro ($r > 10\,\mathrm{kpc}$). In this context, the intermediate meso-scale (parsecs to kiloparsecs) remains one of the least explored regimes in current numerical studies, despite being the critical bridge between halo-scale condensation in cosmological simulations and the inner accretion flows resolved in general-relativistic magneto-hydrodynamical (GR-MHD) black hole simulations. This definition also aligns naturally with our radially concentric SMR design, allowing scale-by-scale diagnostics at a nearly uniform effective resolution within each regime.

The analysis of the results is organised as follows. In Section~\ref{sec:filaments}, we examine the morphology of cold clumps and filamentary structures. In Section~\ref{sec:accretion}, we study the SMBH accretion rate variability in the two turbulence regimes. In Section~\ref{sec:Thermodynamic}, we present radial profiles of the key thermodynamic variables and discuss the implied thermal and dynamical balance of the atmosphere. We then characterise the emergence of multiphase gas using density-temperature phase diagrams (Section~\ref{sec:phases}) and phase mass evolution (Section~\ref{sec:massevol}). Finally, in Section~\ref{sec:multiphase}, we analyse the probability density functions of the condensed multiphase gas.

\subsection{Filamentary structure}\label{sec:filaments}

Turbulence governs the chaotic motions of the hot gas and seeds the density and temperature fluctuations from which the multiphase medium condenses via nonlinear thermal instability \citep{gaspari2013,gaspari2017}. Its strength plays a key role in regulating how gas cools, fragments, and is ultimately accreted onto the SMBH. We begin our analysis by comparing the morphological structure of the filaments and clumps that emerge in the two turbulence regimes \highc\ and \lowc. 

\begin{figure*}
\centering
\includegraphics[width=1\textwidth]{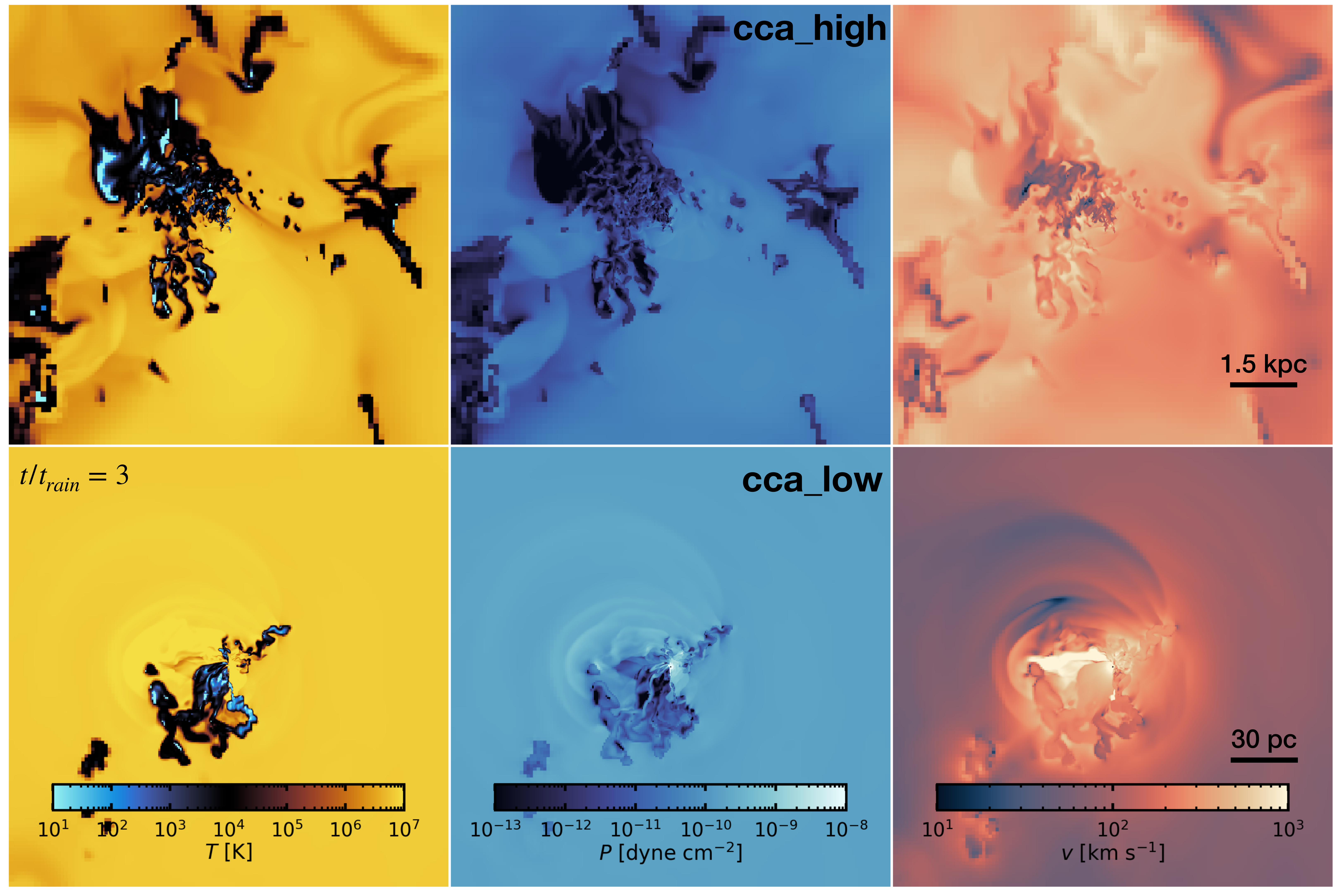}
\caption{Slices of the simulated galaxy group core for the \highc\ (top row) and \lowc\ (bottom row) simulations at $t/t_{\text{rain}}=3$, showing temperature (left), pressure (middle), and velocity magnitude (right). The maps highlight the multiphase structure of the IGrM and the role of turbulence in driving thermodynamic fluctuations in the hot atmosphere, seeding local thermal instability and the condensation of cold gas. Each panel in the top row shows an $8 \,\mathrm{kpc} \times 8 \,\mathrm{kpc}$ region, while each panel in the bottom row shows the central $0.2 \,\mathrm{kpc} \times 0.2 \,\mathrm{kpc}$ region. \label{sliceT}}
\end{figure*}

Figure~\ref{density_proj_highc} shows gas density slices at different times and physical scales in the \highc\ run, illustrating how the gas transitions from an initially smooth medium to a fully developed multiphase structure. The first panel in each row provides a large-scale view of the hot halo showing the full 50 kpc simulation box. At the beginning of the simulation, the hot gas forms a coherent and thermally stable structure, with only mild fluctuations induced by turbulence. As turbulence develops, it stirs the hot atmosphere, generating elongated density fluctuations aligned with the large-scale shear of the flow. Nonlinear compressions induced by turbulence trigger localised catastrophic cooling, generating cold gas filaments and clumps around $t\approx 30$ Myr. By $t/t_{\text{rain}} \approx 1$ (see Figure \ref{density_proj_highc}, first row), the first filamentary condensations appear along regions of converging motions. The turbulent compression locally enhances the density and cooling rate, allowing cold gas to form despite the hot background. The filaments are often aligned with the turbulent eddies and concentrated in the central $\approx 10$ kpc. At later times ($t/t_{\text{rain}} \simeq 2.5$–$3$), the system shows fragmentation of the cold and dense gas driven by the turbulent cascade, which transfers kinetic energy to smaller scales and increases the velocity dispersion in the core to $\sigma_v \approx 230~\mathrm{km~s^{-1}}$. Multiple filaments interact, merge, and break apart under the action of turbulence, cooling, and gravity. The dense structures develop internal denser sub-filaments and knots, forming a tangled network surrounding the central region.

From left to right (Figure \ref{density_proj_highc}), the columns progressively zoom in from the halo scale (tens of kpc, i.e. macro-scale) to the filamentary condensation region ($\approx 1$ kpc, i.e. meso-scale) and finally to the innermost clumps (central parsec, i.e. micro-scale), capturing the end state of the condensation cascade. Towards the galactic centre, this clumpy multiphase complex extends all the way to the SMBH, where thin cold filaments transport gas inward as clouds rain in from kiloparsec scales.

In the \lowc\ simulation (Figure~\ref{density_proj_lowc}), the central $\sim1$~kpc region is shown at $t/t_{\mathrm{rain}} = 1.5$ and $t/t_{\mathrm{rain}} = 3$, zooming in towards the centre of the halo, exhibiting distinct morphology and distribution compared to \highc. The initially smooth inflows condense after $t=7$ Myr, earlier than in the \highc\ case due to the lower turbulent heating-forming a multiphase structure that remains predominantly filamentary rather than clumpy. The resulting morphology is irregular yet more coherent than in \highc, with cold gas circulating in a less disordered pattern. Compared to \highc, the cold phase is more centrally concentrated, being largely confined within the innermost $100$ pc. 

In both simulations, a clumpy rotating structure (see the inset panels in the last panels of Figures~\ref{density_proj_highc} and \ref{density_proj_lowc}) emerges spontaneously at the centre despite the initially non-rotating conditions. Indeed, turbulent eddies induce local vorticity and stochastically broaden the angular momentum distribution in the hot phase (see also \citealt{gaspari2015}); as gas cools and condenses, anisotropic inflows transfer a small residual angular momentum to the collapsing material. Since multiphase inflows do not perfectly cancel out, the remaining angular momentum accumulates in the centre, forming a rotating structure of characteristic radius $\approx 10$~pc in \highc\ and $\approx 5$~pc in \lowc\ composed mainly of cold and molecular gas but with a substantial fraction of warm gas.
Given our resolution and sink treatment, we do not interpret this feature as a fully resolved, rotationally supported accretion disc; rather, it is best viewed as a torus-like, clumpy rotating complex at the micro-scale, potentially analogous to the commonly observed clumpy AGN torus (and/or the outermost reservoir feeding the unresolved inner disc).
This structure is continuously perturbed by infalling filaments; therefore, the inflow remains chaotic and highly time-variable despite the presence of net rotation. These simulations show that the meso-scale (parsecs to kiloparsecs) is not a passive transition region, but an active regime where turbulence and radiative cooling together shape the filamentary network. In this range, cold filaments form, fragment, and interact, establishing the geometry and radial extent of the cold phase before it is accreted towards the central region.

Figure~\ref{sliceT} illustrates the contrasts in the thermodynamic properties by showing slices of temperature (left panel), pressure (middle panel), and velocity magnitude (right panel) across the group core in the \highc\ (top row) and \lowc\ (bottom row) simulations. The temperature map reveals the coexistence of hot ambient gas ($T\approx10^6-10^7$ K) and cooler condensed gas ($T\approx10-100$ K), with a clear multiphase structure within the filaments and clumps, where warm gas ($\approx10^4$ K) surrounds cold molecular cores. The diffuse halo gas instead remains in the $10^6$–$10^7$ K range. The pressure slice reveals thermal pressure contrasts between the cold structures and the volume-filling hot gas, especially in the \highc. Cold structures correspond to regions of lower thermal pressure ($\approx 10^{-13}$–$10^{-12}$ dyne cm$^{-2}$) compared to the surrounding hot medium ($\approx 10^{-10}$–$10^{-9}$ dyne cm$^{-2}$), showing a non-isobaric condensation. Finally, the velocity field shows the strong, chaotic motions that continuously reshape the filamentary network. The velocity maps reveal complex kinematics driven by turbulent motions, spanning velocities from $\sim 10$ to $10^3$~km~s$^{-1}$, with most cold structures being transported and accreted towards the central region. The slices emphasise the contrasting filament distributions, with \highc\ showing extended structures across the central kpc and \lowc\ retaining cold gas confined to the innermost tens of parsecs. An in-depth quantitative analysis of cold clouds and filaments, including their properties and spatial distribution, will be presented in a separate, upcoming work.

\subsection{Accretion rates}\label{sec:accretion}

\begin{figure}
\centering
\includegraphics[width=0.95\columnwidth]{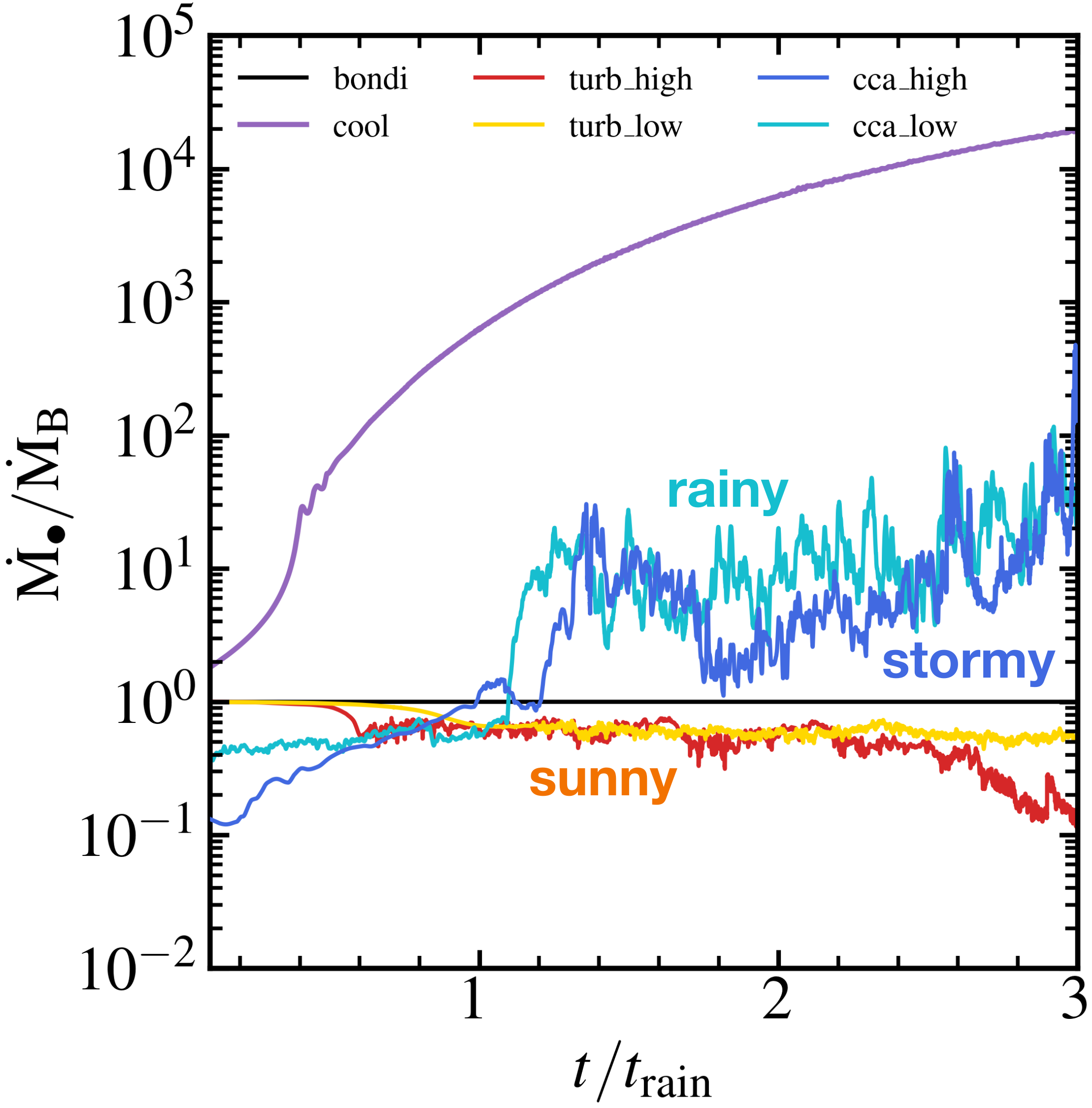} 

\caption{SMBH accretion rate $\dot{M}$ normalised to the Bondi accretion rate $\dot{M_{\rm B}}$ as a function of normalised time $t/t_{\text{rain}}$ for the \highc\ (blue line, stormy weather) and the \lowc\ (cyan line, rainy weather) simulations, compared with the turbulence simulations (sunny weather) \turbohigh\ (red line) and \turbolow\ (yellow line), with the radiative cooling simulation \cool\ (purple line) and with the idealised adiabatic simulation \bondi \ (black line). The CCA phenomenon produces a highly variable accretion rate due to its chaotic nature.}\label{accretionlowc}

\end{figure}

To highlight the differences between these different CCA regimes we analyse the time evolution of the accretion rate in our simulations. Here we provide a brief overview, while a more detailed analysis of gas accretion is presented in the companion paper (B26b). $\dot{M}$ is computed as the mass accreted within the sink region ($r<0.4$~pc) at each simulation timestep. We express the accretion rate in units of the Bondi rate $\dot{M}_{\rm B}$, which provides a natural reference scale for hot mode accretion. Figure~\ref{accretionlowc} shows the normalised accretion rates $\dot{M}/\dot{M_{\rm B}}$ as a function of normalised time $t/t_{\rm rain}$ for the \highc\ (blue, stormy weather) and \lowc\ (cyan, rainy weather) simulations, compared with the adiabatic Bondi run (\bondi, black), the purely turbulent control runs (sunny weather) \turbohigh\ (red) and \turbolow\ (yellow), and the cooling-only simulation \cool\ (purple).

The classical Bondi accretion rate \citep{bondi1952} is given by
\begin{equation}\label{bondieq}
\dot{M}_{\rm B} = \lambda\, 4\pi (G M_\bullet)^2 
\frac{\rho_{\infty}}{c_{{\rm s}, \infty}^{3}},
\end{equation}
where $\lambda=(1/2)^{(\gamma+1)/2(\gamma-1)}[(5-3\gamma)/4]^{-(5-3\gamma)/2(\gamma-1)}$ is a normalisation factor of order unity, $\gamma$ is the gas adiabatic index, $M_\bullet$ is the black hole mass, $\rho$ is the gas density and $c_{\rm s}$ is the gas sound speed (the infinity symbol denotes very large radii). Although this expression is widely employed in numerical and observational studies, often even when the Bondi radius is unresolved \citep[e.g.][]{DiMatteo2005,cattaneo2007,Booth2009,yang2012}, it assumes a steady, homogeneous, adiabatic, and spherically symmetric inflow, which is not found in realistic galactic environments. Therefore, the adiabatic \bondi\ run provides a stable reference for the hot-mode flow but cannot be used to compute a realistic accretion rate. The BH accretion rate has a nearly constant value of $\dot{M}_\bullet\simeq6\times10^{-4}\,\mathrm{M}_\odot\,\mathrm{yr^{-1}}$, in close agreement with the analytic Bondi prediction based on the ambient hot-gas density and temperature (Eq. \ref{bondieq}).

\begin{figure*}
\hspace{-0.5cm}
\includegraphics[width=1.03\textwidth]{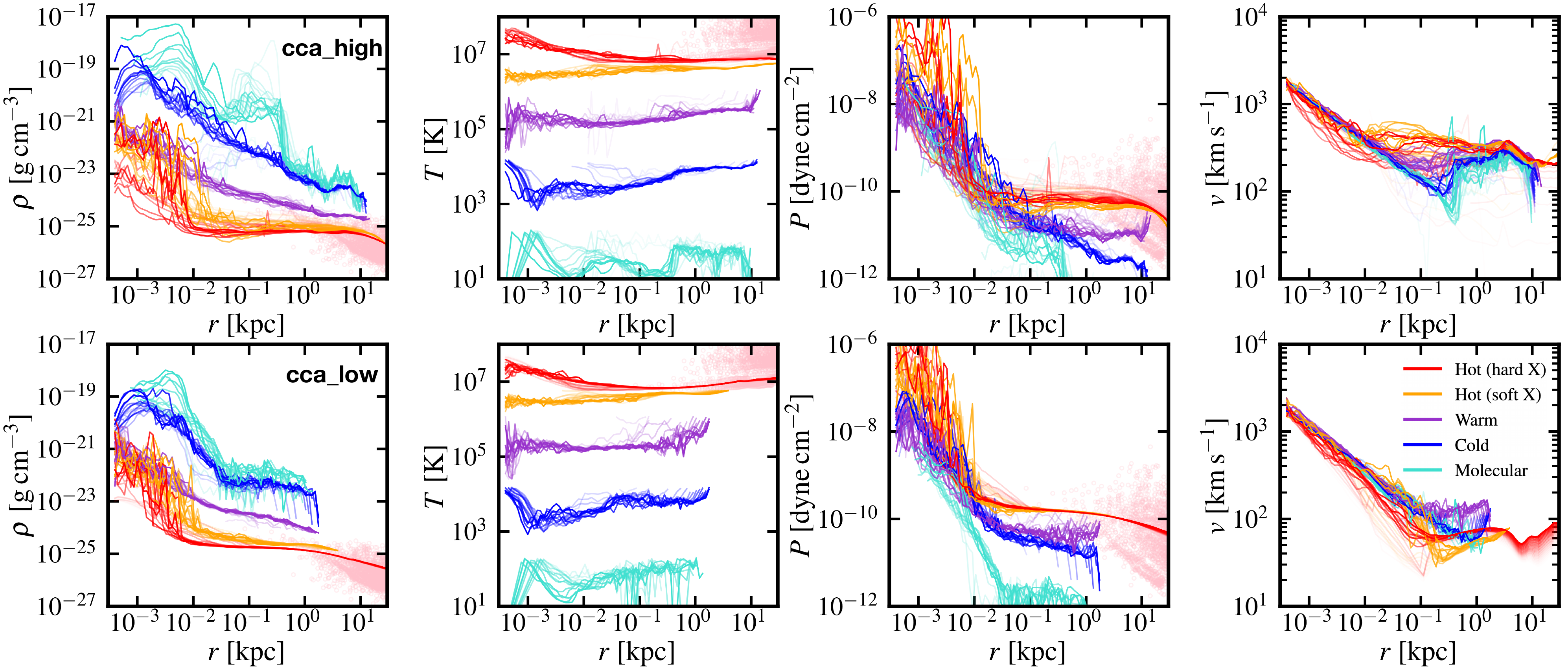}
\vspace{-0.1cm}
\caption{Mass-weighted radial profiles of gas density, temperature, thermal pressure, and velocity magnitude for simulations \highc\ (top row) and \lowc\ (bottom row). The gas is divided into five temperature phases: hot, hard X-ray (red) and soft X-ray (orange), warm (violet), cold (blue) and molecular (cyan, see Table \ref{tab:thermal_phases}). Increasing colour intensity indicates later stages of evolution. Profiles are computed in spherical shells around the group centre out to 25 kpc, showing how the thermodynamic and kinematic structure of the multiphase IGrM evolves in the two CCA regimes. Pink circles show observational density, projected temperature, and pressure profiles from the ACCEPT database \citep{cavagnolo2009}, where applicable.}\label{profiles}
\end{figure*}

The impact of turbulence in the absence of cooling is isolated in the purely turbulent runs. Both \turbolow\ and \turbohigh\ remain firmly in the hot accretion regime, but exhibit systematically sub-Bondi accretion rates. In the \turbolow\ run, characterised by moderate subsonic stirring ($\mathcal{M}\simeq0.1$--$0.2$), the accretion rate fluctuates around $\dot{M}_\bullet \approx 0.6\,\dot{M}_{\rm B}$. Increasing the turbulent velocity dispersion in \turbohigh\ ($\mathcal{M}\simeq0.4$) further suppresses the inflow, yielding $\dot{M}_\bullet \approx 0.5\,\dot{M}_{\rm B}$ and a gradual decline at late times. In these runs, turbulent motions generate vorticity and small rotating eddies that support gas against the accretion, decreasing direct inflow towards the SMBH. In addition, bulk turbulent motions introduce a non-zero relative velocity between the gas and the black hole, reducing the effective accretion rate in a Bondi–Hoyle–like manner \citep{bondi1944}.

The \cool\ simulation shows a strong increase in the accretion rate, from $\approx 1 \ \dot{M}_{\rm B}$ up to $\approx 10^{4} \ \dot{M}_{\rm B}$. This occurs because gas in the central region cools efficiently and accumulates towards the centre, producing accretion rates several orders of magnitude above the Bondi value. This configuration is not realistic, as it neglects perturbations in the medium, which are expected to be present in real astrophysical systems.

Including radiative cooling in a turbulent medium fundamentally changes the accretion regime. Once multiphase condensation sets in, the SMBH transitions from a turbulence-dominated, sunny atmosphere to a CCA mode, corresponding to rainy and stormy weather conditions. The formation of cold filaments and clumps, followed by their inelastic interactions and loss of angular momentum, leads to accretion rates that are strongly super-Bondi. In both \highc\ and \lowc, the accretion rate rises sharply after $t/t_{\rm rain}\sim1$, reaching values one to two orders of magnitude above the Bondi rate. Typical levels of $\dot{M}_\bullet \sim 10$--$100\,\dot{M}_{\rm B}$ are sustained for extended periods, accompanied by strong variability. The accretion becomes highly intermittent, with rapid fluctuations driven by chaotic cloud–cloud collisions and the sporadic infall of cold filaments. Despite their markedly different filamentary morphologies, the \highc\ and \lowc\ runs display qualitatively similar accretion histories once CCA is established, indicating that the presence of multiphase gas dominates the regulation of SMBH fueling. On average, both runs remain strongly super-Bondi, highlighting the inefficiency of Bondi-based prescriptions in capturing accretion in cooling, multiphase halos.

\subsection{Thermodynamic variables radial profiles}\label{sec:Thermodynamic}

To investigate the thermodynamic stratification and time evolution of the multiphase IGrM, we computed mass-weighted radial profiles of density, temperature, pressure, and velocity magnitude for the two fiducial simulations, \highc\ and \lowc\ (Figure~\ref{profiles}). Gas is separated into five temperature phases (see Table \ref{tab:thermal_phases}): hot hard X-ray (red), hot soft X-ray (orange), warm (violet), cold (blue) and molecular (cyan) phases. Each profile represents a mass-weighted average within 3D spherical shells centred on the centre of the simulation, extending out to $\approx25$~kpc. The colour intensity encodes the temporal evolution, from early (faint, $t/t_{\text{rain}}=0$) to late (bright, $t/t_{\text{rain}}=3$) stages. The overlaid ACCEPT profiles (pink circles) show that the simulated hot X-ray phase is broadly consistent with observed IGrM density, temperature, and pressure profiles on kpc scales\footnote{While the simulated temperature profile lies towards the lower part of the ACCEPT distribution, it remains consistent with the observed range of galaxy groups rather than the hotter massive cluster population.}.

In the \highc\ run, the hot plasma fills the entire simulation volume, with densities of $10^{-25}$--$10^{-24}\,\mathrm{g\,cm^{-3}}$ at $r \sim 1~\mathrm{kpc}$ and temperatures of several $10^{7}~\mathrm{K}$. The profiles also reveal a non-negligible amount of gas below $10^{6}~\mathrm{K}$, confined within the innermost $\sim10~\mathrm{kpc}$. As expected, the gas density increases towards colder phases, from the hard X-ray component down to the molecular phase, which reaches $10^{-19}$--$10^{-18}\,\mathrm{g\,cm^{-3}}$ in the central 100~pc. This central region exhibits the strongest variability in the density profiles, as clumps and filaments can be disrupted from their mutual interactions or turbulence effects, while the outer halo remains comparatively stable. The temperature of the hard X-ray phase remains nearly flat beyond the central 100~pc and slightly increases towards the centre. The hot soft X-ray, warm, and cold components show mildly increasing temperatures with radius. The molecular phase instead exhibits small-scale fluctuations with radius but no clear systematic trend, remaining roughly constant on average.

The pressure peaks in the centre and decreases with radius in all gas phases. The hot hard and soft X-ray components show a pronounced central excess of $\approx 10^{-6} \mathrm{dyne \ cm^{-2}}$. The colder phases reach central pressures of $10^{-7}$–$10^{-8} \mathrm{dyne \ cm^{-2}}$ and start to separate beyond $\sim 10$~pc: the warm gas maintains the highest pressure at $\approx10^{-11} \mathrm{dyne \ cm^{-2}}$, followed by the cold phase, while the molecular cold component drops to nearly $10^{-12} \mathrm{dyne \ cm^{-2}}$. This radial decline reflects gravitational stratification and is consistent with approximate local pressure balance between the different phases. Differences between the temperature phases pressure profiles are expected, since the warm, cold, and molecular gas tend to occupy different parts of the multiphase structures, such as mixing layers or filaments centres.

Although the molecular, cold and warm components seem to be thermally under-pressured relative to the ambient hot medium, the hot and cold phases are in approximate pressure equilibrium once non-thermal turbulent pressure \citep[e.g.][]{khatri2016} is included. We defined a total pressure,
\begin{equation}
P_{\rm tot} \equiv P_{\rm th} + P_{\rm nt} \simeq n k_{\rm B} T + \frac{1}{3}\rho\,\sigma_v^2,
\end{equation}
where $k_B$ is the Boltzmann constant and $\sigma_v$ is the gas velocity dispersion. We can simply derive this pressure equilibrium for the molecular component in \highc, as it is the final outcome of the condensation cascade. For the hot phase, adopting typical values of the gas in the central kiloparsec ($n_{\rm hot}\approx 0.1~{\rm cm^{-3}}$ and $T_{\rm hot}\approx 10^7~{\rm K}$, see Section \ref{sec:phases}) yields
\begin{equation}
P_{\rm th, hot} = n_{\rm hot} k_{\rm B} T_{\rm hot} \approx 1.4\times10^{-10}\ {\rm dyne\ cm^{-2}}.
\end{equation}
The corresponding non-thermal term for $\sigma_{\rm hot}\approx 200~{\rm km~s^{-1}}$ is
\begin{equation}
P_{\rm nt, hot} = \frac{1}{3}\rho_{\rm hot}\sigma_{\rm hot}^2
\approx 1.3\times10^{-11}\ {\rm dyne\ cm^{-2}},
\end{equation}
so that $P_{\rm tot, hot}\approx 1.5\times10^{-10}~{\rm dyne\ cm^{-2}}$. In contrast, the cold molecular gas has a much smaller thermal pressure. Using
$n\approx 100$ cm$^{-3}$ and $T\approx 100$ K, we obtain
$P_{\rm th, cold}\approx 10^{-12}$--$10^{-13}\ {\rm dyne\ cm^{-2}}$
(see also the second panel in Figure \ref{sliceT}). The high densities of the molecular gas imply that even modest internal motions provide substantial non-thermal support. Adopting $\mu\simeq 2.3$ for molecular gas and
$\sigma_{\rm cold}\approx 10$ km s$^{-1}$, we obtain
\begin{equation}
P_{\rm nt,cold}= \frac{1}{3}\rho_{\rm cold}\sigma_{\rm cold}^2
\approx 1.3\times10^{-10} {\rm dyne\ cm^{-2}}.
\end{equation} 
Therefore, despite being thermally under-pressured, the cold molecular phase satisfies
$P_{\rm tot, cold}\approx P_{\rm tot, hot}$, implying that the multiphase medium is close to pressure balance when considering the non-thermal pressure component (magnetic fields could further increase this contribution). More broadly, recent SZ observations of AGN-inflated X-ray cavities also suggest that non-thermal pressure components can provide an important contribution to the pressure budget in hot halo atmospheres \citep[e.g.][]{AbdullaCarlstrom2019, Orlowski-SchererHaridas2022}.

Importantly, the degree of thermal under-pressure depends on turbulence strength. In the \highc\ run, strong non-thermal support allows cold clouds to remain dynamically confined while being significantly under-pressured in thermal terms. In contrast, in the \lowc\ case the reduced non-thermal support requires a higher thermal compression of the cold gas, bringing the system closer to an almost isobaric configuration (Figure \ref{sliceT}). The weaker turbulent motions in the hot phase also limit the spatial coherence of cooling compressions, leading to the formation of smaller and more compact cold clouds compared to \highc.

The velocity field is dominated by subsonic turbulent motions on kiloparsec scales, with characteristic speeds of $\approx 300~\mathrm{km\,s^{-1}}$ in the hard and soft X-ray phases and $\approx 200~\mathrm{km\,s^{-1}}$ in the cold and warm gas, with a mild decline towards colder phases, reflecting the decoupling of cooler gas from the hot turbulent eddies and the partial dissipation of turbulent motions during condensation. Towards the innermost parsecs, the flow accelerates in the deep gravitational potential of the SMBH, and the typical velocities rise to $\approx 10^{3}~\mathrm{km\,s^{-1}}$ within the central sub-parsec region.

The \lowc\ run shows broadly similar radial trends but with a more centrally concentrated cold and warm component. The molecular phase reaches high densities in the inner few hundred parsecs, but its density declines more rapidly beyond $r \sim 0.1$~kpc and it is effectively confined within $\sim 1$~kpc. Its temperature remains in the tens to $\approx 100$~K range and the corresponding pressure profile lies systematically below that of \highc\ beyond $r\sim0.01$~kpc, reflecting the smaller radial extent of the coldest gas. The warm and cold phases are likewise more compact, typically extending only out to $\sim 2$–$3$~kpc, while the hot atmosphere dominates at larger radii. Overall, \lowc\ still develops a multiphase core, but the cold and warm gas are less extended and more centrally concentrated than in \highc, consistent with weaker turbulent transport.

\subsection{Phase diagrams}\label{sec:phases}

\begin{figure}
\centering
\includegraphics[width=0.9\columnwidth]{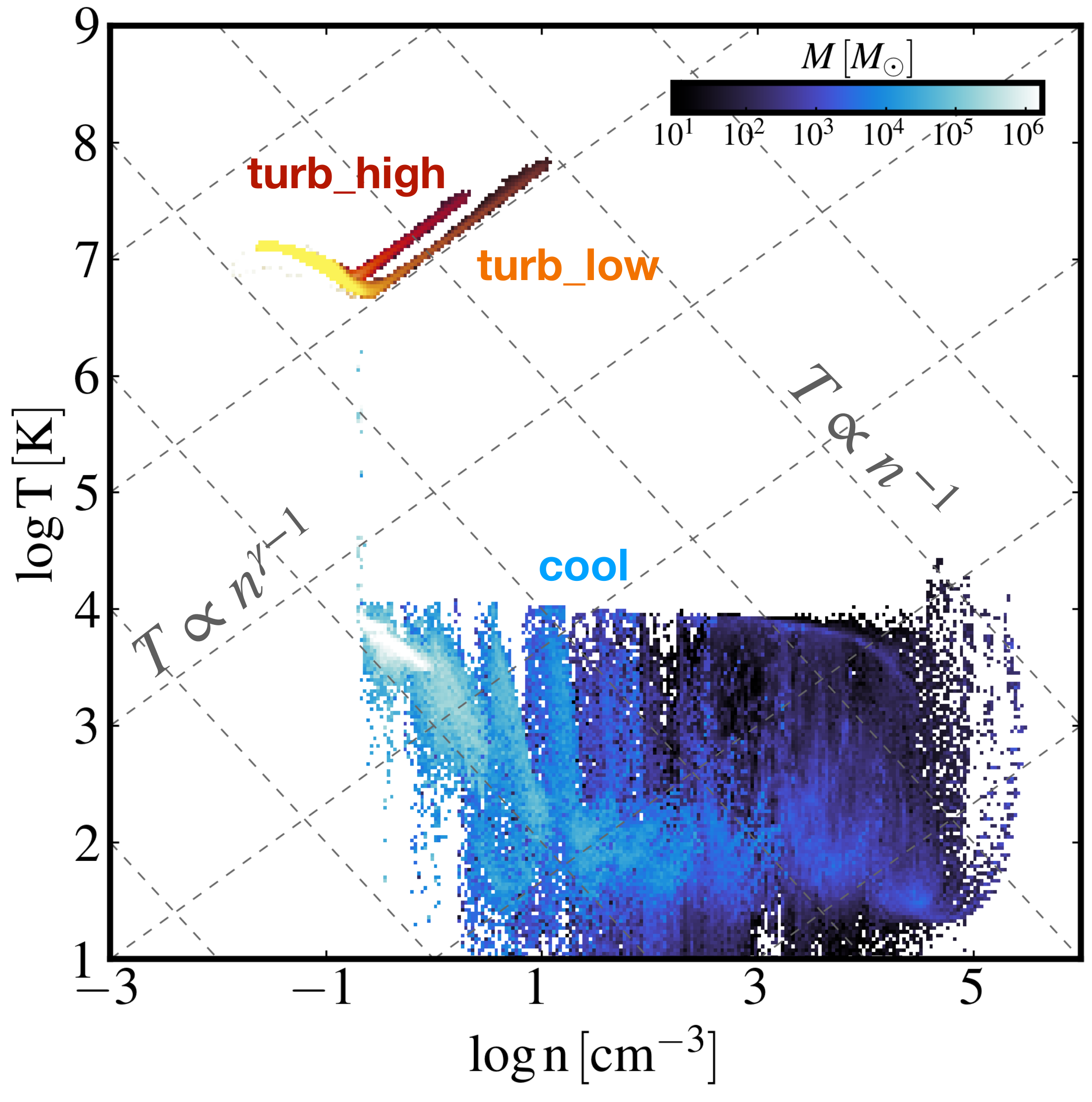} 

\caption{Mass-weighted density-temperature phase diagrams for the \cool\ (blue), \turbohigh\ (red) and \turbolow\ (orange) simulations. Dashed oblique lines indicate adiabatic ($T\propto n^{\gamma-1}=n^{2/3}$) and isobaric ($T\propto n^{-1}$) relations.}\label{phaseplot_comparison}

\end{figure}

To characterise the multiphase nature of the IGrM \citep[as recurrently found in observations, e.g.][]{Sun2009,TemiAmblard2018,Eckert2021,olivaressalome2022,ubertosi2025,temi2026}, we start by comparing the phase diagrams obtained from the different simulations. Figure~\ref{phaseplot_comparison} shows the phase diagrams of the benchmark runs in which radiative cooling is deactivated (\turbohigh\ and \turbolow) and of the run in which turbulent driving is switched off (\cool). The oblique grey dashed lines indicate reference thermodynamic paths. Lines with negative slope represent isobaric evolution ($T \propto n^{-1}$), corresponding to gas cooling or heating while remaining in pressure equilibrium with its surroundings. Lines with positive slope show adiabatic evolution ($T \propto n^{2/3}$), associated with compressions and expansions driven by turbulent motions. These guides help illustrate the dominant thermodynamic behaviour of the gas.

In the turbulence-only simulations, the initial conditions form a stripe at low density. With time, a high-density tail emerges as gas is adiabatically compressed towards the centre, similarly to Bondi accretion. Meanwhile, turbulence causes the initial stripe to widen through repeated adiabatic compressions and expansions. \turbohigh\ maintains gas at systematically higher temperatures, while reaching lower maximum densities with respect to \turbolow. The stronger turbulent motions disrupt the gas in the centre, preventing it from accumulating. The gas remains confined to a hot, low-density branch. As a result, these runs do not produce a realistic multiphase medium, but instead maintain a nearly single-phase atmosphere.

In the \cool\ simulation, the opposite behaviour is observed. Gas undergoes radiative cooling, forming a cooling sphere expanding from the centre in which material rapidly transitions from the hot phase to very cold temperatures, from the stable $T \sim 10^4$ phase down to $10$~K. As a result, cold gas accumulates near the centre and fails to reproduce the extended multiphase structure observed in galaxy groups. A realistic multiphase IGrM does not arise from turbulence or cooling alone, but from their combined action.

\begin{figure*}
\centering

\includegraphics[width=0.91\textwidth]{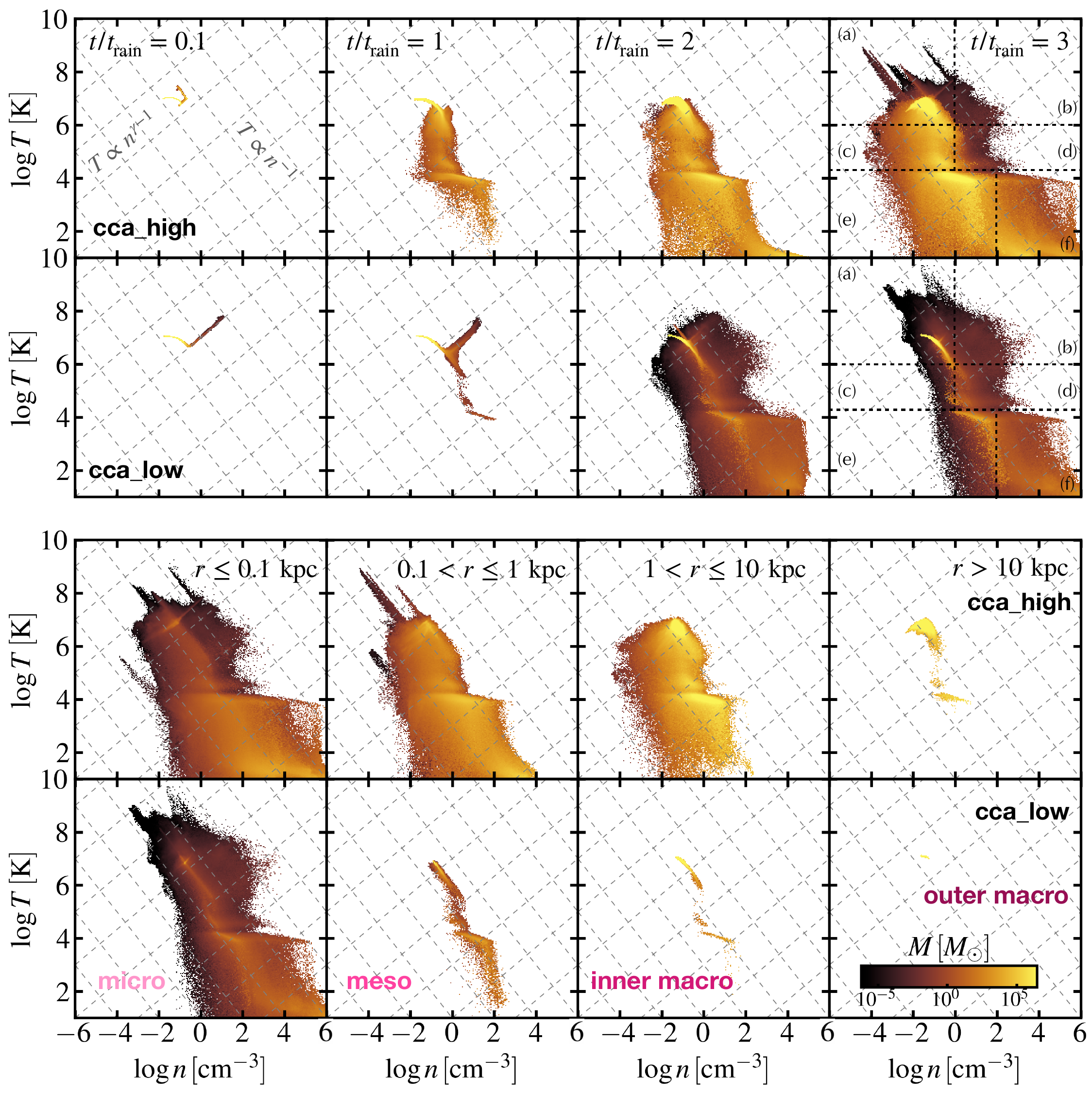} 

\caption{Mass-weighted density--temperature phase diagrams for the \highc\ (rows~1 and~3) and \lowc\ (rows~2 and~4) simulations. 
The first two rows show the time evolution at $t/t_{\mathrm{rain}} = 0.1,\,1,\,2,\,3$, 
while the last two rows show the corresponding distributions at $t/t_{\mathrm{rain}} = 3$ 
in four radial ranges: $0<r\leq0.1$ kpc (micro-scale), $0.1<r\leq1$ kpc (meso-scale), $1<r\leq10$ kpc (inner macro-scale), and $r>10$ kpc (outer macro-scale). Colours denote the total gas mass in each $(n,T)$ bin, from dark brown (low mass -- $10^{-5}$ M$_{\odot}$) to yellow (high mass -- $10^{5}$ M$_{\odot}$). Phases (a-f) are classified by temperature (hot: $T \gtrsim 10^6$ K; warm: $10^4 < T < 10^6$ K; cold: $T < 10^4$ K) and density (diffuse vs. dense, with thresholds at $n = 1\ \mathrm{cm^{-3}}$ for hot/warm and $n = 100\ \mathrm{cm^{-3}}$ for cold). Dashed oblique lines indicate adiabatic ($T\propto n^{2/3}$) and isobaric ($T\propto n^{-1}$) relations in each panel.}\label{phaseplot_4x4}

\end{figure*}

Figure~\ref{phaseplot_4x4} shows the temporal and spatial evolution of the gas density--temperature phase diagram in the \highc\ and \lowc\ fiducial simulations. In the first two rows, the panels illustrate how the phase structure develops in the full simulation volume at increasing times, expressed in units of $t_{\mathrm{rain}}$ 
($t/t_{\mathrm{rain}} = 0,\,1,\,2,$ and~$3$). The last two rows show the gas phase structure at $t/t_{\mathrm{rain}} = 3$, decomposed into the four radial ranges: micro-scale ($r\leq0.1$~kpc), meso-scale ($0.1<r\leq1$~kpc), an inner macro-scale ($1<r\leq10$~kpc), and an outer macro-scale ($r>10$~kpc). In each panel we show dashed oblique lines indicating adiabatic ($T\propto n^{2/3}$) and isobaric ($T\propto n^{-1}$) evolution.

At the beginning of both simulations ($t/t_{\mathrm{rain}} = 0.1$), the gas lies almost entirely in the hot, volume-filling phase with $T \gtrsim 10^6\ \mathrm{K}$ and densities $10^{-2}<n < 1\ \mathrm{cm^{-3}}$, as seen in the ICs profile (Figure \ref{profile_IGM}). The atmosphere is still monophase and in approximate hydrostatic equilibrium. At this stage, the cooling time is much longer than the dynamical time; condensation is negligible and most of the mass remains in the hot turbulent halo. By $t/t_{\text{rain}}=1$, the picture starts to change. The locus in the phase diagram broadens significantly: while a large fraction of the gas remains on the hot branch, a tail of material extends towards higher densities (up to $n\approx10^{2}$ cm$^{-3}$) and lower temperatures. In \highc, the gas spans a broader range of thermodynamical conditions, reaching temperatures as low as $10^2$~K, whereas in \lowc, the cold component is confined to the $10^4$~K phase, indicating a slower progression towards the cold molecular regime. The phase plot at this stage shows gas populating mainly intermediate densities ($n \sim 1{-}10^{2}\ \mathrm{cm^{-3}}$) and temperatures ($T \sim 10^{4{-}5}\ \mathrm{K}$), forming a continuous bridge between the hot ambient medium and the emerging cold phase. 
This is the first signature of thermal instability triggered by the interplay of radiative cooling and turbulent fluctuations. Such overdensities can grow non-linearly when the cooling time becomes sufficiently short relative to the free-fall time, commonly expressed by $t_{\rm cool}/t_{\rm ff}\lesssim 10$ \citep[e.g.][]{field1965,pizzolato2005,gaspari2012,mccourt2012,sharma2012}, thereby forming the chaotic cold rain.
Overdense regions, compressed by turbulent motions, cool more efficiently and move away from the hot phase locus.

At $t/t_{\text{rain}}=2$, a distinct multiphase structure has developed. In addition to the hot background, there is now a well-populated branch extending down to $T \lesssim 10\ \mathrm{K}$ reaching densities up to $n \sim 10^5-10^6\ \mathrm{cm^{-3}}$. Dense clumps and filamentary structures condense at $n \gtrsim 1\ \mathrm{cm^{-3}}$, becoming thermodynamically decoupled from the surrounding medium.

At $t/t_{\text{rain}}=3$, a wider multiphase picture appears. In the \highc\ run, the stronger turbulence produces a highly fragmented and extended phase-space distribution: the gas spans over ten orders of magnitude in density and almost eight in temperature. The cold component reaches down to $T \lesssim 10~\mathrm{K}$ and $n \gtrsim 10^{6}~\mathrm{cm^{-3}}$. Although the hot phase is still dominant, the IGrM is multiphase, signaling a fully developed regime of CCA. The broader extent of the distribution in the \highc\ simulation is a direct consequence of the stronger turbulent motions. Turbulence continuously compresses and expands the gas, driving its evolution along the oblique adiabatic trajectories. These rapid compressions and rarefactions widen the distribution in both density and temperature, preventing the gas from remaining confined to a narrow distribution. As a result, the \highc\ simulation exhibits a significantly broader phase-space structure than \lowc, where weaker turbulence produces more limited density contrasts and a more compact distribution. In contrast, the \lowc\ run remains more compact in phase space. Weaker turbulence limits both density contrasts and the nonlinear efficiency of cooling. Cold gas forms only in the centre (see Figure \ref{density_proj_lowc}) partially, reaching minimum temperatures of $T \sim 10~\mathrm{K}$ and densities up to $n \lesssim 10^{6}~\mathrm{cm^{-3}}$. Most of the mass stays in the hot or warm regime, with a smoother transition between phases and without the sharp multiphase separation seen in the high-turbulence case. In both simulations, and more prominently in \highc, part of the gas evolves along isobaric tracks towards higher temperatures ($T \gtrsim 10^{7}$~K), consistent with heating driven by turbulent dissipation.

The diagram reveals the presence of a rich multiphase medium, we highlight 6 different coexisting components in the forth panels: (a) The diffuse hot phase ($T \gtrsim 10^6$ K, $n < 1\ \mathrm{cm^{-3}}$) represents the volume-filling hot plasma in quasi-hydrostatic equilibrium. (b) The dense hot phase ($T \gtrsim 10^6$ K, $n > 1\ \mathrm{cm^{-3}}$) consists of gas locally compressed by turbulent stirring or shocks, where cooling is about to become effective. (c) The diffuse warm phase ($10^4 < T < 10^6$ K, $n < 1\ \mathrm{cm^{-3}}$) traces the mixing interfaces between the hot atmosphere and condensed filaments. (d) The dense warm phase ($10^4 < T < 10^6$ K, $n > 1\ \mathrm{cm^{-3}}$) marks thermally unstable gas rapidly cooling towards the cold regime. (e) The diffuse cold phase ($T < 10^4$ K, $n < 100\ \mathrm{cm^{-3}}$) forms the extended envelopes of filaments and clumps. (f) The dense cold phase ($T < 10^4$ K, $n > 100\ \mathrm{cm^{-3}}$) corresponds to the molecular gas that can eventually form stars \citep{Ferriere2001}. The presence of ongoing star formation in active AGN environments has been detected also by recent JWST observations \citep[e.g.][]{reefe2025}. In our simulations, cold dense gas is present and could potentially form stars before reaching the central SMBH, provided that the local free-fall time is shorter than the Jeans time (e.g. $t_{\mathrm{ff}} \approx 10~\mathrm{Myr} > t_{\mathrm{jeans}} \approx 1~\mathrm{Myr}$). This scenario is supported by observations revealing star formation within cold filaments located several kiloparsec from the galaxy centre \citep{tremblay2016}. While star formation is not explicitly included in the present work, we plan to investigate its role in future studies. The emergence of the cold and molecular phases reflects the thermodynamic transition driven by density fluctuations induced by turbulence and radiative cooling, leading to the formation of cold clumps and filaments that decouple from the hot halo and sink towards the SMBH enhancing accretion.

The spatial distribution of the gas changes markedly with radius (Figure~\ref{phaseplot_4x4}, third and fourth rows). At the micro-scale, both simulations show the majority of the multiphase gas. This highlights the crucial importance of resolving these small physical scales to correctly study the accretion of gas at the centre of galaxies. Both very hot and very cold gas phases are present in this compact region.

At the meso-scale region the two simulations diverge. In the \highc\ simulation, the enhanced turbulent motions preserve a complex network of cold filaments and clumps embedded within the hot atmosphere. In contrast, \lowc\ still hosts both warm and cold gas, but the IGrM is dominated by more discrete phases — a warm component with $T \approx 10^4$ K and $n \approx 1$–$10^2$ cm$^{-3}$, and a colder, denser component with $n \approx 10^2$ cm$^{-3}$. This suggests that weaker turbulence leads to less efficient mixing and fragmentation.

At the inner macro-scale, the \highc\ simulation retains significant amounts of intermediate-temperature gas ($10^{4.5}$--$10^{6}$~K) and cold gas down to 10 K, whereas the \lowc\ case is mainly composed by the hot phase with a small fraction of warm and $10^4$ K gas. Finally, at the outer macro-scale, both runs are dominated by a tenuous, hot medium ($T \gtrsim 10^{6}$~K), but the \highc\ simulation shows a non negligible amount of cold gas at $T\approx10^4$ K, indicating that turbulence promotes the persistence of multiphase structures even in the outskirts, whereas \lowc\ shows no presence of cold gas. Overall, at the same normalised time, both simulations show a rich multiphase environment but with very different spatial distribution. \highc\ develops a more violent fragmentation and a fully established multiphase structure until $\gtrsim 10$ kpc, while \lowc\ remains more dominated by the hot medium, with smaller and more stable condensations in the central kiloparsec.

\begin{figure}
\centering
\includegraphics[width=\columnwidth]{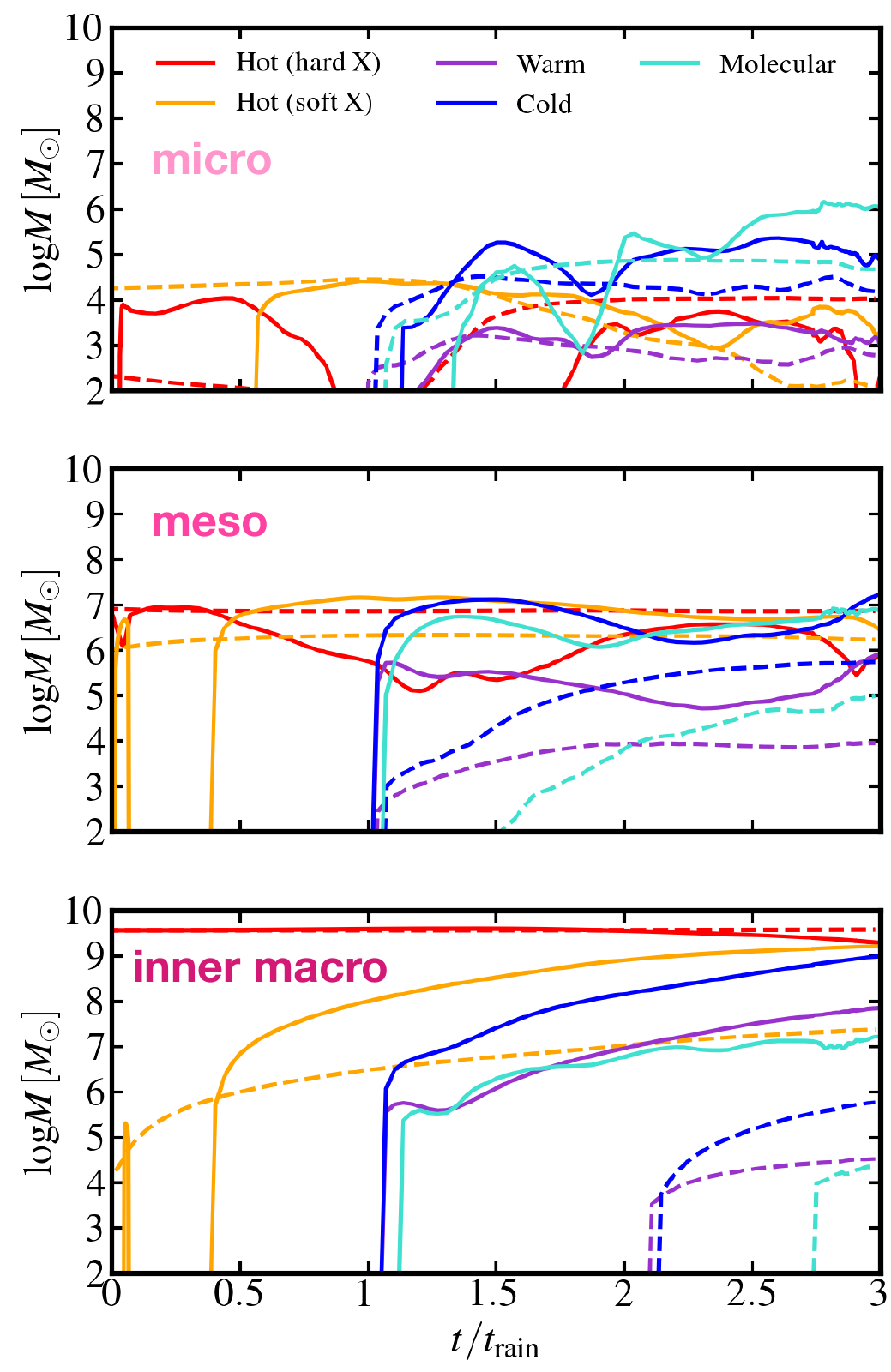} 

\caption{Gas mass for \highc\ (solid) and \lowc\ (dashed) simulations divided in molecular (cyan), cold (blue), warm (violet), soft (orange), and hard (red) X-ray gas and in different scales: macro, meso, inner macro and outer macro.}\label{masstime}

\end{figure}

\subsection{Mass evolution}\label{sec:massevol}

The differences between the two turbulence regimes become clear when examining the amount of gas in the different thermal phases. Figure~\ref{masstime} shows the gas mass of each phase as a function of time at the micro ($r \leq 0.1$~kpc), meso ($0.1 < r \leq 1$~kpc), and inner macro ($1 < r \leq 10$~kpc) scales. The total average mass of each phase for the two simulations is shown in Table \ref{tab:mass_phases_mean} together with the mass ratio. Overall the \highc\ simulation contains larger masses of warm, cold, and molecular gas at all radii. 

At the micro-scale (top panel), all the phases masses in \highc\ vary strongly with time due to the stronger turbulence and the interaction between different clumps and filaments that can heat up or disrupt the cold gas. At $t<t_{\rm rain}$ the medium is dominated by the hot gas in both simulations ($M_{\rm hot} \approx 10^4$ M$_\odot$). After $t/t_{\rm rain}=1$ warm, cold and molecular gas start to appear, with the molecular gas mass reaching $M_{\rm mol} \approx 10^6~\mathrm{M_\odot}$ at $t/t_{\rm rain}=3$. At the same time, the \lowc\ run shows a much smaller molecular mass, $M_{\rm mol} \approx 5\times10^4~\mathrm{M_\odot}$, with a smoother and more gradual increase. A similar behaviour is found for the cold phase, although with lower masses of $M_{\rm cold}\approx10^5$$~\mathrm{M_\odot}$ in \highc\ and $M_{\rm cold}\approx10^4$$~\mathrm{M_\odot}$ in \lowc. In contrast, the warm gas mass is comparable in the two runs, with values around $M_{\rm warm}\approx10^3~\mathrm{M_\odot}$. Even at these small scales, differences between the two regimes are already present.

The meso-scale region (middle panel) marks the first appearance of warm and cold gas in \highc, at $t/t_{\rm rain}=1$. By contrast, in \lowc\ the cold phase first forms at the micro-scale. This highlights a difference in where condensation initially develops, although in both simulations a large fraction of $M_{\rm cold}$ is later found in this region. The cold gas mass in \highc\ ($M_{\rm cold}\approx10^6-10^7$ M$_{\odot}$) remains systematically higher than the warm phase ($M_{\rm warm}\approx10^5-10^6$ M$_{\odot}$) throughout the simulation, without a steady monotonic increase but again with an oscillatory trend showing the strong turbulence and variability of the medium. The molecular gas forms later, after the appearance of the warm and cold phases. This happens at all scales for both simulations and provides insight into the condensation cascade: gas first cools into the warm phase and then rapidly transitions into the colder, more stable phase. As a result, most of the warm gas is confined to the interfaces between hot and cold material within clumps and filaments, as also seen in Figure~\ref{sliceT}. The molecular gas mass in \highc\ remains high, although slightly below the cold gas mass, almost reaching $M_{\rm mol}\approx10^7~\mathrm{M_\odot}$ at $t/t_{\rm rain}=3$. In contrast, in \lowc\ the molecular component is much smaller, appearing only after $t/t_{\rm rain}\approx1.5$ and reaching $M_{\rm mol}\approx10^5~\mathrm{M_\odot}$ at $t/t_{\rm rain}=3$. The hot phase in \highc\ is irregular over time ($M_{\rm hot} \approx 10^5-10^7$ M$_\odot$), whereas in \lowc\ it is almost constant ($M_{\rm hot} \approx 10^7$ M$_\odot$).

\begin{table}
\centering
\caption{
Average gas mass in the different thermal phases for \highc\ and \lowc.
}
\label{tab:mass_phases_mean}
\begin{tabular}{lccc}
\hline
\hline
Phase & \highc\ [M$_\odot$] & \lowc\ [M$_\odot$] & ratio \\
\hline
Hot (hard X)   & $2.6\times10^{9}$ & $3.7\times10^{9}$ & $0.7$ \\
Soft (soft X)   & $1.2\times10^{9}$ & $1.1\times10^{7}$ & $110$ \\
Warm & $5.8\times10^{7}$ & $2.0\times10^{4}$ & $2900$ \\
Cold  & $8.4\times10^{8}$ & $4.3\times10^{5}$ & $2000$ \\
Molecular & $2.4\times10^{7}$ & $8.5\times10^{4}$ & $280$ \\
\hline
\end{tabular}
\tablefoot{
Masses are time-averaged and integrated over the full simulation volume.
}
\end{table}

At the inner macro-scale (bottom panel), the \lowc\ simulation shows almost no warm or cold gas during the early evolution, with these phases starting to grow only after $t/t_{\rm rain}\approx2$. In \highc, at larger distances from the centre, the phase masses display a smoother, less oscillatory evolution and increase steadily with time. As a consequence, the hot gas mass slowly decreases from $\approx5\times 10^9$ M$_{\odot}$ to $\approx10^9$ M$_{\odot}$ as gas cools and transitions into lower-temperature phases. The mass of cold gas in \highc\ strongly increases over time passing from $M_{\rm cold}\approx10^6$ M$_{\odot}$ to $10^9$ M$_{\odot}$, the same happens to the warm phase which increases reaching $M_{\rm warm}\approx10^8$ M$_{\odot}$. The molecular phase is formed soon after also steadily increasing with slight oscillations arriving to $2-3\times10^7$ M$_{\odot}$ by the end of the simulation.

Despite the \highc\ simulation developing a substantially larger cold and molecular gas reservoir (Table~\ref{tab:mass_phases_mean}), the SMBH accretion rate remains comparable, and in some phases slightly lower, to the one measured in the \lowc\ run (see Figure \ref{accretionlowc}). This indicates that BH feeding is not directly set by the total mass of cold gas present, but rather by the efficiency with which cold structures can lose angular momentum and be transported towards the centre. This is consistent with high-resolution ALMA observations showing no clear correlation between $\lesssim 100$ pc molecular gas mass and AGN activity tracers \citep{ElfordDavis2024}. In the stormy regime, a significant fraction of the cold gas remains stored in extended, rotationally supported and dynamically stirred structures, while only a limited portion effectively participates in accretion. Conversely, in the rainy regime, cold gas forms less abundantly but is accreted more efficiently once it condenses, leading to comparable inflow rates despite the much smaller cold gas reservoir.

\subsection{Multiphase density structure across scales}\label{sec:multiphase}

\begin{figure*}
\centering
\includegraphics[width=0.9\textwidth]{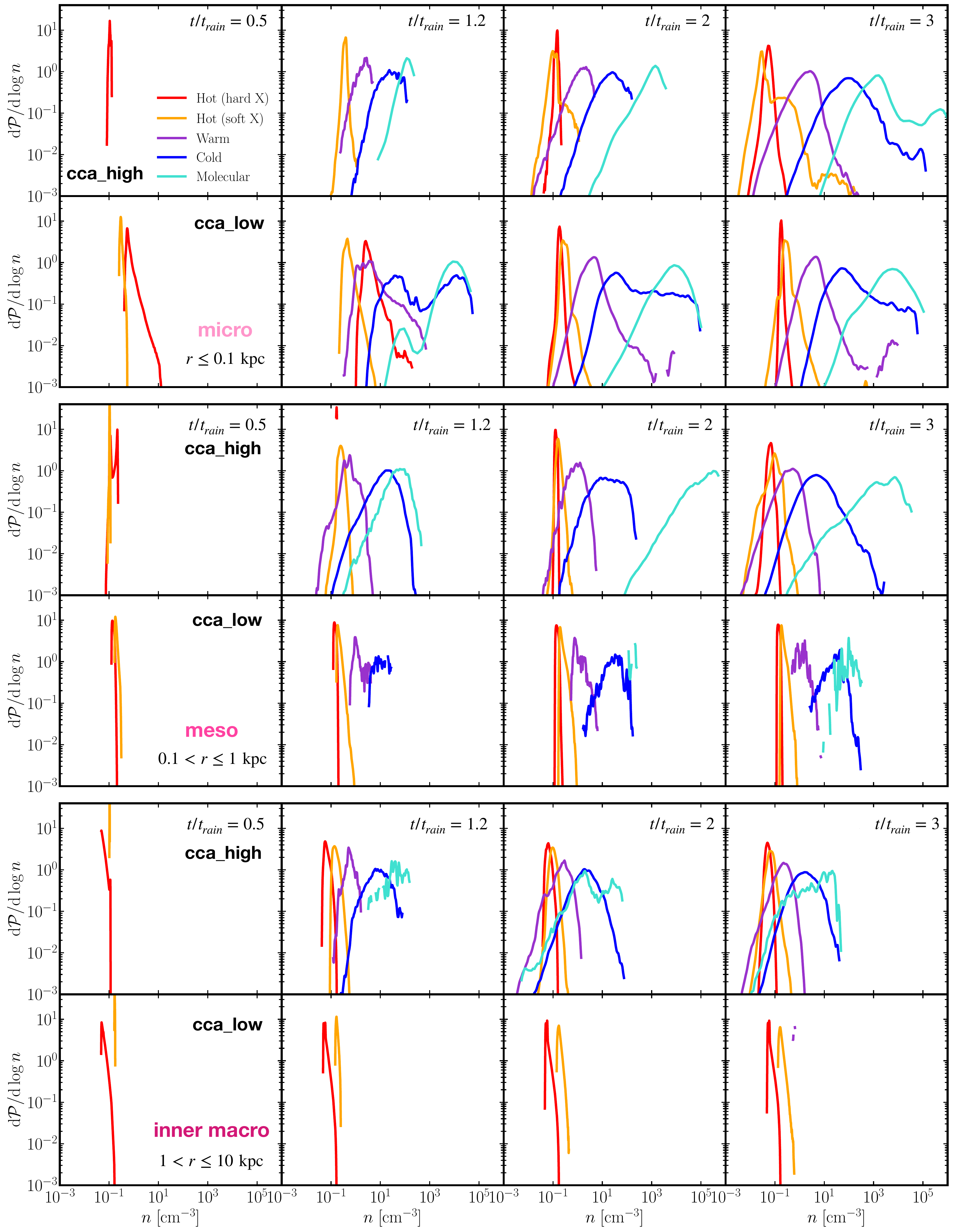}

\caption{Probability density functions (PDFs) of gas density measured in three radial shells, spanning the micro, meso, and inner macro-scales.
The top two rows show the evolution within the inner $r \leq 0.1~\mathrm{kpc}$ region (micro-scale), the middle two rows correspond to $0.1 < r \leq 1~\mathrm{kpc}$ (meso-scale), and the bottom two rows to $1 < r \leq 10~\mathrm{kpc}$ (inner macro-scale).
For each radial shell, the upper row displays the \highc\ (strong turbulence) run, while the lower row shows the \lowc\ (weak turbulence) case.
Different columns represent successive snapshots, with times normalised to the rain timescale $t/t_{\mathrm{rain}}$.
Coloured curves trace the various thermal phases: hot hard X-ray (red), hot soft X-ray (orange), warm (violet), cold (blue), and molecular (cyan).}\label{PDFmicro}

\end{figure*}

In Figure~\ref{PDFmicro} we present the time evolution of the mass-weighted gas density probability distribution functions (PDFs) for different thermal phases, measured in three radial shells spanning the micro, meso, and inner macro regimes. For each shell we select all gas cells within the specified radial range and bin them uniformly in $\log n$. The PDFs are mass-weighted: in each density bin $i$ we compute the mass fraction
$p_i \equiv \frac{1}{M_{\rm shell}}\sum_{j\in i} m_j$,
where $m_j$ is the mass of cell $j$ and $M_{\rm shell}=\sum_j m_j$ is the total gas mass in the shell. The corresponding differential PDF is $d\mathcal{P}/d\log n \simeq p_i/\Delta\log n$, with $\Delta\log n$ the bin width. By construction, $\sum_i p_i = 1$ (equivalently $\int (d\mathcal{P}/d\log n)\,d\log n = 1$), so the integral over any $\log n$ interval yields the corresponding mass fraction. We adopt 150 bins spanning $\log n=-3$ to 6 and quantify each phase via the mass-weighted mean $\mu_{\log n}$ and dispersion $\sigma_{\log n}$ (and the skewness $\mathcal{S}_{\log n}$ where quoted below).

The innermost 100~pc region (micro-scale), where most of the cooling and condensation occur, is shown in the top two panels of Figure \ref{PDFmicro}.
In \highc, the initially hot and nearly homogeneous core ($n \sim 0.1~\mathrm{cm^{-3}}$) exhibits a narrow, nearly lognormal density PDF, as expected for subsonic, pressure-supported turbulence in the hot phase. After $t/t_{\mathrm{rain}}\approx1$ a cold molecular component with \(\mu_{\log n}\simeq 2.0\) dex and a relatively small dispersion (\(\sigma_{\log n}\simeq0.2\) dex) has emerged, with strongly negative skewness, indicating that most of the molecular mass sits in a well-defined high-density peak with only a shallow tail towards lower densities. Cold and warm gas populate smaller densities (\(\mu_{\log n}\simeq1.4\) dex and \(0.3\) dex, respectively) with moderate dispersions (\(\sigma_{\log n}\sim0.3\)–0.5 dex) and near-symmetric skewness, tracing filaments and clumps in the intermediate stage of the condensation cascade. At \(t/t_{\mathrm{rain}}\simeq2\) there is a slight broadening of the PDFs in all phases, while maintaining a similar average and distribution. At later times (\(t/t_{\mathrm{rain}}\simeq3\)) the molecular PDF broadens a lot, the mean shifts to \(\mu_{\log n}\simeq4.3\) dex with \(\sigma_{\log n}\simeq1.6\) dex, and skewness becomes mildly positive, which could signal the recurrent formation and disruption of dense clumps. The hot X-ray phases, in turn, become more dilute (soft X-ray \(\mu_{\log n}\) drops from \(\simeq-0.4\) dex to \(\simeq-1.2\) dex) and develop positive skewness, which is consistent with cooling interfaces embedded in this volume-filling medium, which indicates the hot gas from which the condensation cascade is generated. Also the distributions of the warm and cold phases have broadened over time, spanning intermediate densities between the hot and molecular gas. Taken together, the broadening and skewness evolution likely reflect intermittent compressions and mixing layers at hot--warm--cold interfaces; in the companion paper (B26b) we will corroborate this interpretation using scale-dependent velocity statistics and kinematic CCA diagnostics (such as k-plots; \citealt{gaspari2018}).

Also in the \lowc\ simulation  the gas becomes strongly multiphase. By \(t/t_{\mathrm{rain}}\sim1\), a molecular component appears with \(\mu_{\log n}\simeq1.2\) dex and moderate dispersion, while cold and warm phases occupy intermediate densities (\(\mu_{\log n}\simeq1.6\) dex and \(0.4\) dex). The hot gas remains relatively dense and positively skewed. By \(t/t_{\mathrm{rain}}\simeq3\), the coldest gas reaches high densities (\(\mu_{\log n}\simeq5.3\) dex for the molecular phase), but its PDF stays narrower than in \highc\ and the hot components retain higher mean densities and very positive skewness, pointing to rare cooling sites embedded in a largely smooth hot core. In \highc, the hot gas PDFs on average extend to lower densities (\(\lesssim 10^{-2}\,\mathrm{cm^{-3}}\)), reflecting the stronger compression–rarefaction cycle driven by turbulence in this regime.

At meso-scales (middle panels, $0.1 < r \leq 1~\mathrm{kpc}$), both simulations still exhibit multiphase structures, but with different intensities. In \highc, the hot gas becomes progressively more dilute and structured as condensation proceeds. From $t/t_{\mathrm{rain}}\simeq1$ to 3, the soft X-ray phase shifts from $\mu_{\log n}\simeq -0.6$ dex to $\simeq -1.1$ dex, with dispersions increasing from $\sigma_{\log n}\sim0.1$ dex to $\sim0.2$ dex and skewness evolving from mildly positive to mildly negative values. The hard X-ray component follows a similar trend, maintaining $\mu_{\log n}\simeq -0.8$ to $-1.2$ dex with small dispersions ($\sigma_{\log n}\sim0.04$--0.11 dex) and fluctuating skewness, indicative of intermittent turbulent compressions and rarefactions. This evolution reflects the continuous depletion of the densest hot gas as it cools and feeds the colder phases. The warm and cold components populate intermediate densities throughout this radial range, with typical means $\mu_{\log n}\simeq -0.4$--0.1 dex and $\mu_{\log n}\simeq 0.7$--1.3 dex respectively, and moderately broad PDFs ($\sigma_{\log n}\sim0.3$--0.5 dex). Their skewness is generally mildly negative or near zero. The molecular phase is the most intermittent: its PDF alternates between relatively narrow and very broad shapes, with $\mu_{\log n}$ ranging from $\sim0.6$ to $\sim3$ dex and dispersions up to $\sigma_{\log n}\sim0.6$ dex, tracing episodic formation and destruction of dense clumps driven by CCA.

In \lowc, the meso-scale atmosphere remains significantly more stable. Both the soft and hard X-ray phases retain nearly constant mean densities over time, with $\mu_{\log n}\simeq -0.63$ to $-0.70$ dex for the soft X-ray gas and $\mu_{\log n}\simeq -0.83$ to $-0.85$ dex for the hard X-ray component. Dispersions remain small ($\sigma_{\log n}\lesssim0.1$ dex), and skewness stays systematically positive, particularly for the soft X-ray phase ($\mathcal{S}_{\log n}\sim1$--1.6), indicating localised, non-runaway cooling perturbations embedded in an otherwise smooth hot medium. The hard X-ray PDFs remain close to lognormal at all times. Cold and warm phases are present but evolve weakly: their mean densities and dispersions change little with time, and their PDFs remain relatively narrow compared to \highc. The molecular component appears later than in the strong-turbulence case, becoming appreciable only after $t/t_{\mathrm{rain}}\approx2$, with $\mu_{\log n}\sim2$ dex and modest dispersions. Overall, between $0.1$ and $1~\mathrm{kpc}$, \highc\ develops increasingly dilute, negatively skewed hot gas and a broad, intermittent cold component, whereas \lowc\ maintains denser, positively skewed hot phases and a more weakly evolving multiphase structure, reflecting the reduced efficiency of turbulence-driven condensation at meso-scales.

At the inner macro-scale (bottom panels, $1<r\leq10~\mathrm{kpc}$) the thermodynamic evolution of the two simulations is very different. In \highc, the hot soft and hard X phases have an evolution similar to the other scales. Cold phases are already present from \(t/t_{\mathrm{rain}}\simeq1\): a diffuse warm component with \(\mu_{\log n}\sim -0.3\) dex to \(-0.1\) dex and an optical component with \(\mu_{\log n}\sim0.2\)–0.9 dex (dispersions \(\sigma_{\log n}\sim0.3\)–0.5 dex) trace low-level condensation and mixing at meso–macro transition scales. A molecular component appears intermittently with \(\mu_{\log n}\sim0.6\)–1.6 dex and broad PDFs, signaling sporadic transport of dense clumps out to a few kiloparsecs.

In \lowc, the same radial range remains much closer to a single-phase hot atmosphere for most of the evolution. From \(t/t_{\mathrm{rain}}=0\) to 2, only the hot phases are present. Only by \(t/t_{\mathrm{rain}}\simeq3\) a weak multiphase component emerge: faint molecular, cold, and warm phases appear with \(\mu_{\log n}\sim1.2\) dex, $1.4$ dex, and \(-0.1\) dex, respectively, and relatively narrow dispersions, indicating that the condensed gas remains a minor, centrally sourced component rather than an extended filament network.

Overall, between 1 and 10~kpc, \highc\ sustains intermittent cooling and partial multiphase mixing, with cold and warm gas present – albeit in small amounts – throughout the inner halo. In \lowc, the hot phase stays denser and more stable, and multiphase structure at these radii only appears late and remains weak, confirming that most condensation is confined to the central kiloparsec.

\begin{figure*}
\centering
\includegraphics[width=\textwidth]{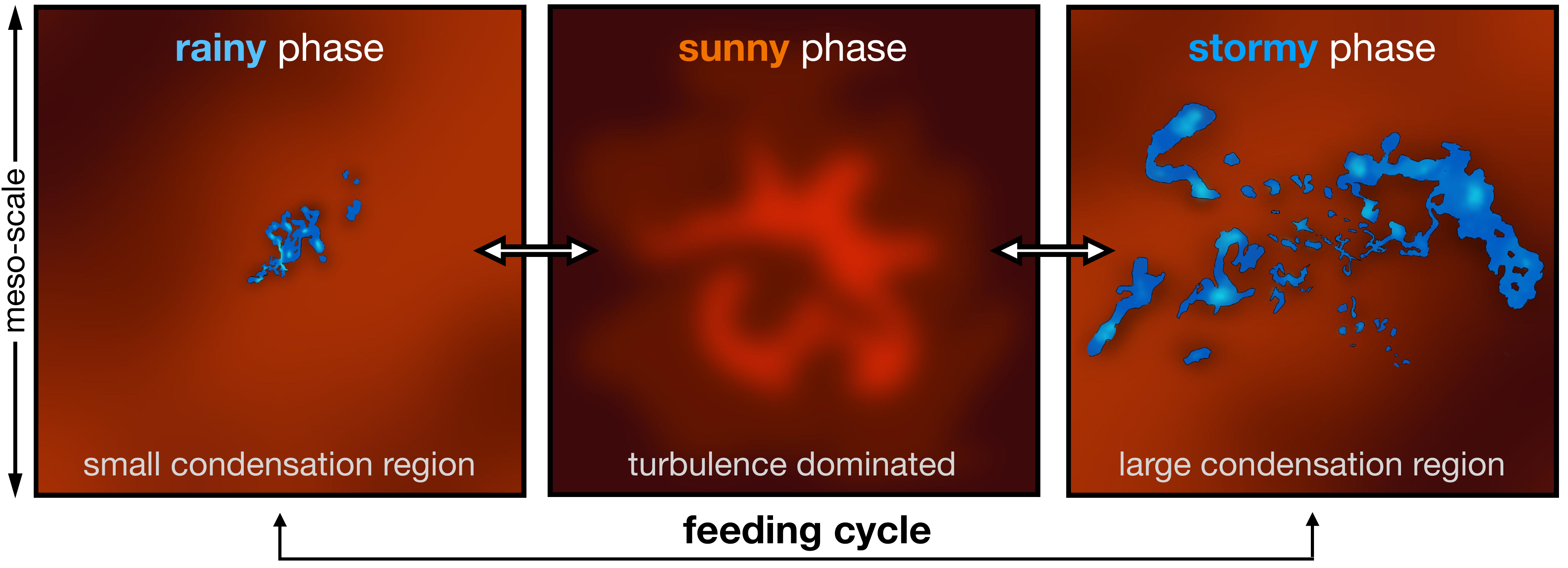}

\caption{Cartoon scheme illustrating a possible evolutionary sequence driven by the interplay between turbulence and gas condensation. Rainy phase (\lowc): in a low-turbulence state, cold clumps and filaments are concentrated near the nucleus, leading to coherent accretion and a centrally confined cold phase. Sunny phase (\turbolow/\turbohigh): increased turbulent stirring redistributes the cold gas and temporarily suppresses further condensation, producing a more diffuse, extended atmosphere. Stormy phase (\highc): in a strongly stirred cooling atmosphere, multiphase filaments reform over a broader radial region, before progressively collapsing towards the nucleus as turbulence weakens, ultimately returning the system to the initial, centrally concentrated state.}\label{scheme}

\end{figure*}

In the outer macro-scale halo (panel not shown, $10<r\leq50~\mathrm{kpc}$), both simulations are trivially dominated by diffuse hot plasma. In \highc, only at late times ($t/t_{\mathrm{rain}}\simeq2-3$) faint warm and cold components appear, with \(\langle\log n\rangle\lesssim0\).

In \lowc, the outer region remains almost perfectly single-phase. At all epochs, only the hard X-ray gas is present, with \(\langle\log n\rangle\simeq -1.54\) to \(-1.49\) and small dispersions. No warm, cold, or molecular phases develop in this region, indicating that condensation and turbulent mixing are effectively confined to smaller radii.

The density PDFs provide a compact, quantitative summary of the multiphase condensation cascade across scales. 
In the strong-turbulence case (\highc), the hot phases progressively develop broader, increasingly asymmetric distributions, while the cold and molecular components become highly intermittent, with extended high-density tails that appear early and persist from the micro- and meso-scale out to the inner macro region. 
This behaviour is consistent with a turbulent condensation cascade in which mixing layers and repeated compressions continuously seed nonlinear overdensities, leading to recurrent formation and disruption of dense clumps. In contrast, the weak-turbulence case (\lowc) shows systematically narrower PDFs outside the central region, with hot phases remaining closer to a quasi-stationary atmosphere and cold gas largely confined to the inner scales. The resulting multiphase medium is therefore more centrally concentrated and less stochastic at meso--macro radii, indicating a reduced efficiency of turbulence-driven fragmentation and transport. Overall, the PDF evolution links the morphological picture (filaments and clumps) and the phase-diagram view into a single scale-resolved statistical framework: stronger stirring broadens the thermodynamic distributions and sustains multiphase structure to larger radii, while weaker stirring favours a more stable hot halo with localised condensation.

\section{An evolutionary link between the two turbulence regimes}\label{sec:evol}

Although \texttt{cca\_high} and \texttt{cca\_low} are designed as two fixed-turbulence experiments, they do not represent intrinsically different classes of group atmospheres. Instead, they can be interpreted as two limiting manifestations of the same underlying CCA process along a feedback--feeding cycle, in which the level of stirring and the efficacy of radiative condensation vary in time. In this context, the `BH weather' terminology provides a compact way to connect the phenomenology seen in the simulations to physical regimes: (i) a sunny state, in which turbulence and heating dominate over cooling so that the atmosphere remains largely hot and condensation is suppressed or strongly reduced, (ii) a stormy CCA state, in which multiphase condensation is active over an extended region and produces a filament-rich network at meso to macro radii, and (iii) a rainy CCA state, in which condensation is still present but is more centrally concentrated, leading to a compact, clumpy cold core at micro to inner-meso radii.

Our results naturally map onto this picture. The two fiducial runs bracket the outcome of CCA under different stirring levels, while preserving the same stratified halo and cooling physics. The strongly stirred case (\texttt{cca\_high}) delays the onset of raining (with $t_{\rm rain}=30$~Myr versus $7$~Myr in \texttt{cca\_low}) and sustains a broader, more radially extended multiphase medium: cold and warm structures persist to larger radii, the density PDFs broaden markedly beyond the central region, and the morphology is dominated by a tangled filamentary network that reaches kiloparsec scales. Conversely, the weak-turbulence case (\texttt{cca\_low}) exhibits faster condensation but a more localised precipitation pattern: the hot halo outside the central region remains closer to a quasi-stationary configuration with comparatively narrower PDFs, while cold gas is largely confined to the inner tens of parsecs and organises into a compact, clumpy multiphase core. These differences show that the same halo can host qualitatively distinct `weather' outcomes depending on how turbulence redistributes, mixes, and compresses the gas during the condensation cascade.

A simple evolutionary interpretation, schematically summarised in Figure~\ref{scheme}, is as follows. In a low-stirring phase (rainy-like), condensation is concentrated towards the nucleus and the cold phase is spatially confined, which can in principle promote stronger central inflow and trigger an AGN response (winds or jets) in a full feedback model. The injected energy and momentum would then heat the atmosphere and raise the turbulent velocity dispersion, potentially shifting the system towards a hotter, turbulence-dominated state in which multiphase condensation is temporarily reduced (sunny-like). As the driving weakens and turbulence decays, radiative cooling progressively regains importance and condensation can restart over a broader radial range, producing a filament-rich, extended precipitation configuration (stormy-like) reminiscent of \texttt{cca\_high}. With further decay of stirring and continued cooling, the condensation region can contract again and cold structures can collapse back towards the nucleus, returning the system towards a more centrally concentrated, rainy-like configuration closer to \texttt{cca\_low}. In this narrative, stormy and rainy represent two CCA flavours regulated primarily by how turbulence distributes multiphase gas in radius, rather than two distinct accretion modes. 

The characteristic timescales in our simulations support the plausibility of such a cycle. 
For example, the onset of raining differs by a factor $\sim 4$ between the two endpoints ($t_{\rm rain}=30$~Myr in \texttt{cca\_high} versus 7~Myr in \texttt{cca\_low}): stronger stirring increases mixing and non-thermal support, leading to a longer raining time and a more extended multiphase structure once condensation develops, whereas weaker stirring produces earlier raining and a compact multiphase core. While our setups impose constant turbulent forcing and therefore cannot demonstrate temporal transitions within a single run, they show that modest changes in turbulence amplitude are sufficient to move the same group-scale halo between an extended, filament-rich CCA configuration and a centrally concentrated, compact multiphase state. The companion paper (B26b) will further connect the weather classification to kinematic CCA diagnostics and time-domain variability, providing an independent corroboration of the physical picture proposed here.

Differences in weather states may also arise from differences in halo mass. Lower-mass groups, with shallower gravitational potentials, may be more susceptible to AGN-driven uplift and turbulent stirring, allowing cold gas to be displaced more efficiently and favouring extended stormy configurations. Conversely, more massive groups may confine the condensed phase more effectively, promoting faster recycling towards the nucleus and producing compact, centrally concentrated rainy-like reservoirs. Thus, the distinction between stormy and rainy weather may reflect both temporal variability within individual systems and a halo-mass dependence in the likelihood of different CCA states.

The companion jet-regulated {\sc BlackHoleWeather} simulations including self-consistent AGN feedback (cf.~C26a,b) suggest that, in those simulations, a fully sunny state is rarely achieved. Even when AGN-driven turbulence and heating suppress condensation in the central region, multiphase gas can continue to survive and condense at larger radii, where jet-driven uplift, mixing, and compression around the outflow cones can locally trigger thermal instability and promote cold gas formation. This naturally gives rise to an intermediate `cloudy' regime, in which the inner atmosphere is hotter and more turbulence-dominated while cold clouds persist at the meso-scale. Complementary {\sc BlackHoleWeather} simulations including SMBH spin evolution and jet precession (cf.~P26a,b) instead show transitions from rainy or stormy configurations towards a sunny phase at the meso-scale, with sub-Bondi accretion rates ($10^{-4}-10^{-5}$ M$_\odot$ yr$^{-1}$). A possible interpretation is that jet precession distributes the feedback energy more isotropically across the central atmosphere, increasing the efficiency of volumetric heating and suppressing condensation more globally. This suggests that the long-term weather cycle may depend sensitively not only on the feedback power, but also on the geometry and coupling of AGN energy injection.

\section{Comparison with previous works}\label{sec:synthesis}

Our setup is intentionally based on the stratified-halo CCA framework introduced in \citet{gaspari2013,gaspari2017} which showed that in a turbulently stirred atmosphere radiative cooling can trigger a nonlinear top-down multiphase condensation cascade: warm filaments and colder clouds form out of the hot phase, inherit its turbulent kinematics, and rain towards the centre where chaotic cloud collisions promote angular-momentum cancellation and strongly time-variable SMBH feeding. The phenomenology we recover is largely consistent with these key CCA expectations: extended multiphase structures arise once cooling becomes competitive with turbulent stirring with correlated thermo-kinematics, the accretion rate is recurrently boosted by orders of magnitude relative to the classical hot-mode baseline, and the fueling remains highly bursty.
At the same time, our results advance the CCA framework in a regime that is particularly relevant to link the meso-micro scales under different BH weather conditions. By explicitly contrasting two subsonic turbulence levels, we show that changes in the turbulent weather (stormy vs. rainy) are sufficient to move the same group-scale halo between an extended, filament-rich CCA precipitation state and a more centrally concentrated CCA configuration with a compact, clumpy inner structure. Advancing here the dynamic range in one calculation further allows us to follow how large-scale filaments progressively fragment into smaller clumps and streams across the three main phases with improved spatial and temporal resolution. This provides an important controlled reference baseline for the forthcoming {\sc BlackHoleWeather} extensions that include jets, spin, and dust, and for more direct confrontation with multi-wavelength constraints.

Other numerical works studied SMBH accretion with high-resolution simulations, albeit with significantly different conditions. \citet{ChoPrather2024} developed a multizone GR-MHD framework that self-consistently bridges the Bondi scale down to the event horizon. While their simulations achieve a global steady-state solution over several decades in radius, the absence of radiative cooling restricts the accretion flow to a purely hot mode. As a result, despite realistic outer boundary conditions and strong magnetic fields, the mass accretion rate remains strongly suppressed ($\sim$\,1\% of the Bondi rate), with no contribution from cold or multiphase accretion. \citet{Guo_2023} simulated an M87-like elliptical galaxy with radiative cooling and distributed heating down to the SMBH micro-scales (with an aggressive first-order flux correction), finding that despite cold discs and chaotic phases seeded by small initial perturbations, the innermost accretion flow remains hot-mode dominated with suppressed accretion rates. \citet{Guo_2024} extended this to MHD, showing that magnetic fields create filamentary cold inflows, enhance angular momentum transport, and boost accretion by an order of magnitude while launching powerful polar outflows. At variance, we use 3D hydrodynamic simulations of a hot intragroup halo with driven subsonic turbulence that models realistic astrophysical turbulence injection and weather \citep[e.g.][]{wittor2020}, as found by multi-wavelength observations. Rather than seeding the cold phase through small decaying initial perturbations, our turbulence continuously drives multiphase condensation, allowing us to track multiple realistic raining episodes and isolate how turbulence shapes the multiphase structure.

\section{Summary and conclusions}\label{conc}

In this work, we analysed the process of gas accretion onto SMBHs in group-scale halos, focusing on how turbulence regulates the condensation, structure, and feeding of multiphase gas. We performed a set of high-resolution ($\Delta x_{\min}\simeq0.1$~pc) 3D hydrodynamical simulations with \athenapk, including radiative cooling and driven, subsonic turbulence in a hot, stratified atmosphere representative of the IGrM. Our two fiducial runs explore contrasting turbulence regimes, a strongly stirred halo (\highc) and a weakly stirred one (\lowc), and are evolved through recurrent raining episodes. Thanks to the SMR, we follow the condensation cascade from tens of kpc across the crucial meso-scale (parsecs to kiloparsecs) down to the inner $0.1$~kpc, resolving well within the Bondi radius and explicitly tracking how cold clouds and filaments feed the SMBH. 
Our main results can be summarised as follows.
\vspace{-0.09cm}
\begin{itemize}
\setlength{\itemsep}{0.3em}
\setlength{\parskip}{0pt}
\setlength{\parsep}{0pt}
\item[(i)]  
In both turbulence regimes, radiative cooling in a stratified, turbulently stirred intragroup medium triggers nonlinear multiphase condensation and rain, giving rise to CCA. Turbulence primarily regulates the radial distribution and morphology of the condensed phase: strong stirring (\highc) delays the onset of raining and sustains extended, filamentary structures reaching kpc scales (stormy weather), whereas weaker stirring (\lowc) yields a more centrally concentrated, clumpy multiphase core, with most cold gas confined within the inner tens of parsecs (rainy weather).

\item[(ii)] 
The meso-scale (parsecs to kiloparsecs) is an active, controlling regime rather than a passive bridge. In this range, filaments repeatedly form out of the hot phase, fragment into clumps and streams, and interact through collisions and shearing layers before being funneled towards the inner region. 
This sustained, multiscale condensation cascade couples halo precipitation to micro-scale inflow, in part mediated by a clumpy, rotating nuclear structure reminiscent of an AGN torus.

\item[(iii)] 
We directly follow cold clumps and filaments down to sub-pc radii and measure the SMBH accretion rate at the sink. In both \highc\ and \lowc, the accretion rate is recurrently boosted by $\sim 10$--$100$ relative to the classical hot-mode (Bondi-like) baseline, with stochastic excursions of up to $\sim 2$ dex, reflecting the intrinsically chaotic nature of CCA.

\item[(iv)] 
Although the strongly stirred case develops a much larger cold and molecular reservoir than the weakly stirred case, the accretion rates remain comparable. This demonstrates that SMBH feeding is not set by the total condensed mass, but by how efficiently the condensed structures couple to the central inflow once they form.

\item[(v)] 
Cooling in a turbulent halo produces a strongly multiphase medium spanning $\sim 10$ orders of magnitude in density and $\sim 8$ in temperature. In \highc, stronger stirring broadens thermodynamic distributions and sustains warm and cold gas to larger radii, yielding markedly broader density PDFs beyond the nucleus. In \lowc, density contrasts outside the centre remain smaller, the outer halo stays closer to a classical hot atmosphere, and the density PDFs beyond $\sim 100$ pc are narrower and more time-stationary. Across phases, incorporating turbulent support reduces apparent thermal pressure imbalances, bringing the multiphase medium closer to local pressure balance in terms of $P_{\rm tot}=P_{\rm th}+P_{\rm nt}$.

\end{itemize}

Taken together, these results show that the explored turbulence regimes are best interpreted as different `BH weather' realisations within the same CCA process. Stronger stirring delays condensation and favours an extended, filament-rich precipitation pattern (a longer, hot turbulence-dominated sunny stage in a full cycle, followed by stormy CCA), whereas weaker stirring promotes earlier condensation and a compact, centrally concentrated multiphase core (rainy CCA). In this first paper we therefore focus on the thermodynamic and morphological manifestation of multiphase condensation across scales, quantified through phase structure, PDFs, and mass budgets. The companion paper \citep{Barbani2026b} builds directly on the same simulations to quantify the condensation criteria (via CCA diagnostics such as $\mathcal{C}$-ratios and k-plots) and to connect the weather states to time-domain inflow variability and its statistical signatures.

Overall, the simulations presented here provide a new step towards a unified, multiscale theory of AGN feeding in group atmospheres. By combining stratified intragroup initial conditions, driven subsonic turbulence, radiative cooling, and GPU-enabled dynamic range, we follow in a single calculation the full CCA cascade from halo rain at tens of kpc down to $\sim 0.1$~pc. In doing so, we corroborate the core CCA framework established by \citet{gaspari2013,gaspari2017,GaspariTombesi2020} and extend it by quantifying how distinct `BH weather' regimes imprint on multiphase morphology and thermodynamics across the meso-scale. This controlled reference baseline sets the stage for forthcoming {\sc BlackHoleWeather} studies that will incorporate additional physics and confront the predicted diagnostics with multi-wavelength observations across a broad range of environments and scales.\\

\begin{acknowledgements}
We thank the anonymous referee for constructive comments and suggestions. The BHW authors acknowledge key funding support from the European Research Council (ERC) under the European Union's Horizon Europe research and innovation programme (Consolidator Grant BlackHoleWeather, No.~101086804; PI: Gaspari). Views and opinions expressed are, however, those of the author(s) only and do not necessarily reflect those of the European Union or the European Research Council Executive Agency; neither the European Union nor the granting authority can be held responsible for them.
We acknowledge ISCRA for awarding this project access to the LEONARDO supercomputer, owned by the EuroHPC Joint Undertaking, hosted by CINECA (Italy). The numerical work was in part supported by the NASA High-End Computing (HEC) Program through the NASA Advanced Supercomputing (NAS) Division at Ames Research Center. VO acknowledges support from the DICYT ESO-Chile Comite Mixto PS 1757, and Fondecyt Regular 1251702. FMM acknowledges support from the Next Generation EU funds within the PNRR, Mission 4 - Education and Research, Component 2 - From Research to Business (M4C2), Investment Line 3.1 - Strengthening and creation of Research Infrastructures (project IR0000034 “STILES”). MF acknowledges funding by the Deutsche Forschungsgemeinschaft (DFG, German Research Foundation) under Germany's Excellence Strategy -- EXC 2121 ``Quantum Universe'' --  390833306.
PT acknowledges support from NASA NNH22ZDA001N Astrophysics Data and Analysis Program under award 24-ADAP24- 0011.
RS acknowledges funding from the CAS-ANID grant No.~220016.
The authors thank Dong-Woo Kim for providing the observational data used in Figure~\ref{profile_IGM}, Sara Sandoni for providing the illustration used in Figure \ref{scheme}, and Philipp Grete for support with the \athenapk\ code. We thank the organisers and participants of the following conferences for the stimulating discussions that helped improve this work: `BlackHoleWeather I' (Sexten, ITA), `Modelling of Multiphase Astrophysical Media' (Ringberg, GER), `Multi-phase, Multi-temperature, and Complex' (Olbia, ITA).

\end{acknowledgements}

   \bibliographystyle{aa}
   \bibliography{biblio.bib}

\end{document}